%% file: Manuscript.tex
\documentclass[11pt]{article}

\usepackage{amsmath,amsfonts,amssymb,amsthm,color,hyperref,url,geometry,setspace,natbib,setspace,ulem}
\hypersetup{
    colorlinks=true,
    linkcolor=blue,
    filecolor=magenta,      
    urlcolor=cyan,
    citecolor=blue,
}
\usepackage{graphicx,booktabs,multirow}
\usepackage{algorithm,algcompatible}
\usepackage{subcaption}
\usepackage{soul}
\setstcolor{red}
\usepackage[noend]{algpseudocode}
\graphicspath{{./figures/}}
\numberwithin{equation}{section}
\input{./math-commands.tex}
\title{Simultaneous Best Subset Selection and Dimension Reduction via Primal-Dual Iterations}

\usepackage[affil-it]{authblk} 
\usepackage{etoolbox}
\usepackage{lmodern}

\makeatletter
\patchcmd{\@maketitle}{\LARGE \@title}{\fontsize{16}{19.2}\selectfont\@title}{}{}
\makeatother

\author[1]{Canhong Wen}
\author[1]{Ruipeng Dong}
\author[1]{Xueqin Wang}
\author[2]{Weiyu Li}
\author[3]{Heping Zhang\thanks{Correspondence: Heping Zhang (heping.zhang@yale.edu)}}
\affil[1]{International Institute of Finance, School of Management, University of Science and Technology of China, Anhui, China}
\affil[2]{School of Gifted Young, University of Science and Technology of China, Anhui, China}
\affil[3]{School of Public Health, Yale University, New Haven, CT}

\date{}

\begin{document}
  \maketitle

  \begin{abstract}
   Sparse reduced rank regression is an essential statistical learning method. In the contemporary literature, estimation is typically formulated as a nonconvex optimization that often yields to a local optimum in numerical computation. Yet, their theoretical analysis is always centered on the global optimum, resulting in a discrepancy between the statistical guarantee and the numerical computation. In this research, we offer a new algorithm to address the problem and establish an almost optimal rate for the algorithmic solution. We also demonstrate that the algorithm achieves the estimation with a polynomial number of iterations. In addition, we present a generalized information criterion to simultaneously ensure the consistency of support set recovery and rank estimation. Under the proposed criterion, we show that our algorithm can achieve the oracle reduced rank estimation with a significant probability. The numerical studies and an application in the ovarian cancer genetic data demonstrate the effectiveness and scalability of our approach.
    \smallskip 

    \noindent{\textbf{Keywords}: High-dimensional data; Multivariate response; Primal-dual active selection algorithm; Reduced rank regression; Variable selection.} 
  \end{abstract}

  \input{introduction.tex}

  \input{model.tex}

  \input{method.tex}

  \input{theory.tex}

  \input{simulation.tex}

  \input{real-data.tex}

  \input{discussion.tex}

  \bibliographystyle{plainnat}
  \bibliography{mycita}

  \input{appendix.tex}

\end{document}

%% file: math-commands.tex
\newcommand{\norm}[1]{\|#1\|}
\newcommand{\wh}[1]{\widehat{#1}}
\newcommand{\wt}[1]{\widetilde{#1}}
\newcommand{\rank}[1]{\text{rank}(#1)}
\newcommand{\diag}[1]{\text{diag}\{#1\}}
\newcommand{\tr}[1]{\text{tr}\{#1\}}
\newcommand{\abs}[1]{\vert #1 \vert}

\newcommand{\op}{\text{op}}
\newcommand{\ora}{\text{ora}}
\newcommand{\GIC}{\text{GIC}}

\def\trans{^{\rm T}}
\def\strans{^{*\rm T}}

\def\code#1{\texttt{#1}}

\newtheorem{theorem}{Theorem}
\newtheorem{lemma}{Lemma}
\newtheorem{condition}{Condition}
\newtheorem{definition}{Definition}

\newtheorem{corollary}{Corollary}

\newcommand{\bDelta}{\boldsymbol{\Delta}}

\newcommand{\bSigma}{\boldsymbol{\Sigma}}
\newcommand{\bGamma}{\boldsymbol{\Gamma}}

\newcommand{\bXi}{\boldsymbol{\Xi}}

\newcommand{\balpha}{\boldsymbol{\alpha}}

\newcommand{\bgamma}{\boldsymbol{\gamma}}

\def\C{{\bf C}}
\def\X{{\bf X}}
\def\Y{{\bf Y}}

\def\V{{\bf V}}
\def\D{{\bf D}}
\def\I{{\bf I}}
\def\E{{\bf E}}

\def\A{{\bf A}}
\def\B{{\bf B}}

\def\Q{{\bf Q}} 
\def\T{{\bf T}}

\def\bP{{\bf P}}

\def\mbR{\mathbb{R}}

\def\x{\boldsymbol{x}}
\def\y{\boldsymbol{y}}
\def\e{\boldsymbol{e}}
\def\c{\boldsymbol{c}}
\def\u{\boldsymbol{u}}
\def\v{\boldsymbol{v}}

\def\a{\boldsymbol{a}}
\def\b{\boldsymbol{b}}

\def\0{\boldsymbol{0}}
\def\1{\boldsymbol{1}}

%% file: introduction.tex
\section{Introduction}

Sparse reduced rank regression is an indispensable tool for many contemporary applications, such as  bi-clustering \citep{lee2010biclustering,zhong2022biclustering}, microeconomics \citep{uematsu2019high}, pattern recognition \citep{zheng2014multi} and others. For the estimate of underlying coefficients, the classical framework for the pursuit of the sparse reduced rank regression has been ``optimization + regularization,'' and many approaches using the regularization technique have been proposed in recent years, including \citet{chen2012sparse,bunea2012joint,she2017selective,uematsu2019sofar,ma2020adaptive}. In the aforementioned works, estimation is characterized as a nonconvex optimization subject to sparsity and low-rank constraints. Scientists then establish a good statistical convergence rate, sometimes even the minimax optimal rate, by concentrating on the optimization's global optimum.

Unfortunately, the nonconvexity that causes the gap between the statistical guarantee and the numerical computation means that the numerical methods in the existing literature usually only achieve a local optimum. There is some published work that makes an attempt to close this gap by imposing the KKT-type condition and the sparsity assumption on the numerical solution (see, for example, \citet{zhang2012general,fan2014asymptotic,chen2022}). The question of whether the obtained numerical solution holds up to these assumptions remains open. It's also not easy to tell if the solution actually matches the goal in their theory, as there's more than one local optimum. As such, it only rephrases the original question and does nothing to bridge the theoretical gap that the numerical algorithm suffers from. Although the sparse reduced rank regression has been numerically solved, the theoretical guarantee for these solutions remains unknown.

The computational complexity of massive data sets is another obstacle. In order to address the original optimization problem, researchers break it down into a series of unit-rank estimations using regularization (see, for example, \citet{chen2012reduced,ahn2015sparse,mishra2017sequential}) and then use alternating coordinate descent to find solutions. \citet{he2018boosted} and \citet{chen2022} have recently argued that alternating coordinate descent is still slow when there are a lot of predictors. Therefore, they employ the stagewise algorithm to decompose sparse matrices. However, the stagewise approach mainly relies on an empirical choice of step size. A slow computation will arise from too small a step size, whereas a reckless estimation will follow from too large a step size. Importantly, the iteration complexity for achieving the statistical convergence rate is unanswered by either the alternating coordinate descent or the stagewise algorithm. Concerning large-scale computing, this is an extremely important issue.



Taking inspiration from the primal-dual algorithm, we offer a new approach to estimating the parameters in sparse reduced rank regressions that fixes the aforementioned problems. In order to find the best subset of covariates, the suggested method iteratively updates the primal and dual variables. Every iteration of the algorithm, which can be interpreted as a local second-order algorithm, implements a typical reduced rank estimation based on the given covariates. Consequently, it benefits from a rapid computation without having to decide on a suitable step size in advance. Additionally, the statistical guarantee for the numerical solution is established. Our contributions are multifold. First, the nearly minimax optimal rate for the algorithmic solution is determined, bridging the gap between theoretical analysis and numerical computation. Second, our approach can produce the nearly optimal estimation in a polynomial iteration complexity, which permits a fast computing in practice. Third, a generalized information criteria (GIC) is introduced to simultaneously ensure the consistency of support set recovery and rank estimation. Last but not least, owing to the proposed GIC, the algorithmic solution achieves the oracle reduced rank estimation with a significant probability.

The remainder of the paper is structured as follows: The proposed method, together with its justification and derivatives, are described in Section \ref{sec:method}. The theoretical results for the proposed method are provided in Section \ref{sec:theory}. Numerical studies are presented in Section \ref{sec:simulation} to demonstrate the advantages of the proposed method. In Section \ref{sec:real}, we show how this can be used in genetics. Future work is discussed in  Section \ref{sec:dis}. All technical proofs are provided in the appendix.

\bigskip 

\noindent\textbf{Notations.} We use the boldface uppercase or lowercase to denote the matrix or column vector, respectively. For an arbitrary vector $\a=[a_1,\dots,a_p]\trans\in\mbR^p$, define the norm $\norm{\a}_{q}=(\sum_{i=1}^p\abs{a_i}^q)^{1/q}$ with $0\leq q \leq +\infty$. For an arbitrary matrix $\A=(a_{ij})\in\mbR^{m\times n}$, we denote the $i$th row of $\A$ by $\a_i\in\mbR^n$, and the $j$th column of $\A$ by $\breve{\a}_j\in\mbR^m$. That is $\A=[\a_1,\dots,\a_m]\trans=[\breve{\a}_1,\dots,\breve{\a}_n]\in\mbR^{m\times n}$. Throughout the paper, we use $\norm{\A}_F = \sqrt{\sum_{i,j}\abs{a_{ij}}^2}$ and $\norm{\A}_{2,0} = \sum_{i=1}^{m}I(\norm{\a_i}_2^2\neq 0)$ to denote the Frobenius norm and the number of nonzero rows of $\A$, respectively. Moreover, denote the operator norm of $\A$ by $\norm{\A}_{\op}$. For an arbitrary index set $I$, $\A_{I\cdot}$ is the submatrix consisting of the rows of $\A$ indexed by $I$. Similarly, $\A_{\cdot I}$ is the submatrix consisting of the columns indexed by $I$ in the original matrix. In addition, $\abs{I}$ denotes the cardinality of $I$. 
%

%% file: model.tex
\section{Methodology and Algorithm}\label{sec:method}

\subsection{Model setting}

Denote the response vector and the covariate vector by $\y\in\mbR^q$ and $\x\in\mbR^p$, respectively. We consider the reduced-rank regression model as follows, 
\begin{align*}
    \y = \C\strans\x + \e,
\end{align*}
where $\e\in\mbR^q$ is the noise vector and $\C^*\in\mbR^{p\times q}$ is the coefficient matrix with $\rank{\C^*}=r^*$. Assume that a random sample of $n$ i.i.d. observations $\{(\x_i,\y_i)\}_{i=1}^n$ is drawn from the above model, then we have 
\begin{align*}
    \Y = \X\C^* + \E,
\end{align*}
where $\Y = \left[\y_1,\dots,\y_n\right]\trans\in\mbR^{n\times q}$, $\X = \left[\x_1,\dots,\x_n\right]\trans\in\mbR^{n\times p}$ and $\E = \left[\e_1,\dots,\e_n\right]\trans\in\mbR^{n\times q}$ denote the matrices of stacked response, covariate and noise vectors, respectively. We denote the $i$th row and the $j$th column of the coefficient matrix by $\c_i^*\in\mbR^q$ and $\breve{\c}_{j}^*\in\mbR^p$, respectively. Moreover, we assume the $\ell_2$ norm of the each column of $\X$ is normalized to $\sqrt{n}$ without loss of generality. 

In this paper, we consider the row-sparsity structure and denote the number of the nonzero rows of $\C^*$ by $s^*$. Then let the index set of nonzero rows be 
\begin{align*}
    A^* = \left\{1 \leq i \leq p: \norm{\c_{i}^*}_2 \neq 0\right\}. 
\end{align*}
Our main target is to recover the row-sparse and low-rank coefficient matrix and its corresponding support set $A^*$. 

Formally, with the user-specified rank $r$ and sparsity $s$, we consider the best subset selection problem in reduced-rank regression as follows 
\begin{align}\label{opt:1}
    \underset{\C\in\mbR^{p\times q}}{\min}~ (2n)^{-1}\norm{\Y - \X\C}_F^2, ~ \text{s.t.}~ \rank{\C} \leq r, ~ \norm{\C}_{2,0} \leq s,
\end{align}
where the $\ell_0$-norm constraint not only enables the variable selection but also reduces the bias in the estimation. Henceforth, we call the problem~\eqref{opt:1} as \underline{M}ulti-\underline{r}esponse \underline{Be}st \underline{S}ubset \underline{S}election (MrBeSS) problem and its minimum as MrBeSS estimator.

%% file: method.tex
\subsection{MrBeSS with fixed rank and sparsity level}\label{sec:derivation}

We next consider the following optimization problem:
\begin{align}\label{eqn:C:fixed}
    \underset{\C\in\mbR^{p\times q}}{\min}~ (2n)^{-1}\norm{\Y - \X\C}_F^2, ~ \text{s.t.}~ \rank{\C} = r, ~ \norm{\C}_{2,0} = s.
\end{align}
The two constraints are intertwined, making the optimization extremely nonconvex to solve. To address this issue, we disentangle these two restrictions through a reparameterization of $\C$, i.e., $\C = \B\V\trans$, with $\B\in \mbR^{p\times r}$ and $\V\in\mbR^{q\times r}$ satisfying $\V\trans\V = \I_r$. Then \eqref{eqn:C:fixed} can be rewritten as 
\begin{align}\label{opt:2}
    \underset{\B\in\mbR^{p\times r},\V\in\mbR^{q\times r}}{\min}~ 
    (2n)^{-1}\norm{\Y - \X\B\V\trans}_{F}^2,  ~ \text{s.t.}~ \V\trans\V = \I_r,  \norm{\B}_{2,0}=s.
\end{align}
Note that the solution to the optimization problem \eqref{opt:2} is not unique. For example, suppose $(\wh{\B}, \wh{\V})$ is a solution of \eqref{opt:2}, then $(\wt{\B}, \wt{\V})$ is also a solution of \eqref{opt:2}, if there is an orthogonal matrix $\Q$ such that $\wt{\B} = \wh{\B} \Q$ and $\wt{\V} = \wh{\V} \Q$. Nonetheless, the fact that $\wh{\C} :=\wh{\B} {\wh{\V}}^\top= \wt{\B} {\wt{\V}}^\top$ assures that both solutions give the same estimation of $\C^*$.

Initialize $k=0$ and assume that we have an initial support set $A^k$. Therefore, restricted in the $A^k$, the optimization \eqref{opt:2} is an ordinary reduced rank regression that has a closed form \citep{Reinsel1998}. It motivates us to update right singular vectors, where the $j$-th column of $\V^{k+1}$ is the $j$-th eigenvector of 
\begin{align*}
  \Y\trans\X_{\cdot A^k}(\X_{\cdot A^k}\trans\X_{\cdot A^k})^{-1}\X_{\cdot A^k}\trans\Y,
\end{align*}
and the eigenvectors are listed in descending order by eigenvalues. Then with the given $\V^{k+1}$, the optimization \eqref{opt:2} is simplified as 
\begin{align}
  \underset{\B\in\mbR^{p\times r}}{\min}~ 
    (2n)^{-1}\norm{\Y\V^{k+1} - \X\B}_{F}^2,  ~ \text{s.t.}~ \norm{\B}_{2,0}=s,
    \label{opt:3}
\end{align}
which is a typical multi-response regression. For the best subset selection of predictors, we consider the following primal-dual formulation 
\begin{align}\label{eq:kkt:approx}
    \left\{
        \begin{array}{l}
          \B_{A^{k}\cdot} = (\X_{\cdot A^{k}}\trans\X_{\cdot A^{k}})^{-1}\X_{\cdot A^{k}}\trans\Y\V^{k+1} \\
          \B_{I^{k}\cdot} = \0_{(p-s)\times r}
        \end{array}
    \right.
    ~\text{and} ~
    \left\{
        \begin{array}{l}
          \bGamma_{A^{k}\cdot} = \0_{s\times r} \\
          \bGamma_{I^{k}\cdot} = \X_{\cdot I^{k}}\trans(\Y\V^{k+1} - \X\B^{k+1})/n
        \end{array}
    \right.,
\end{align}
where $\B_{A^k\cdot},\B_{I^{k}\cdot},\bGamma_{A^{k}\cdot}$ and $\bGamma_{I^{k}\cdot}$ are submatrices of $\B^{k+1}$ and $\bGamma^{k+1}$, respectively, and they consist of rows indexed by corresponding support sets. Let $\b_j^{k+1}$ and $\bgamma_j^{k+1}$ be the $j$-th rows of $\B^{k+1}$ and $\bGamma^{k+1}$, respectively.
With the given $\V^{k+1}$, the optimization \eqref{opt:3} is reduced to a group Lasso, thereby, which is similar to \citet{zhu2020polynomial}. $\delta_j^{k+1}\triangleq \norm{\b_j^{k+1}+ \bgamma_j^{k+1}}_2^2$ can be seen as a measure of the importance of the $j$-th predictor, and $\delta_j^{k+1}$ is equal to the difference of the optimal value of the loss function in \eqref{opt:3} when we add/knockout the $j$-th covariate for the current support set. 
A more significant $j$-th covariate is indicated by a bigger value of $\delta_j^{k+1}$. But the proposed method includes the low rank restriction that distinguishes the problem from  \citet{zhu2020polynomial}.

Based on the above formulation, we develop an algorithm to iteratively update the triple $(\B^k,\bGamma^k,\V^k)$ until the convergence and the outline of the proposed algorithm is presented in Algorithm \ref{alg:mrbess}. Here we need to specify an initial active set $A^0$. A naive argument is to implement the screening for the rows of $\X\trans\Y\V^0$ that selects the indices of top-$s$ rows in term of the $\ell_2$ norm, where $\V^0$ is the top-$r$ right singular vectors of $\Y$. In details, there is $A^0=\{j: \norm{\wt{\x}_j\trans\Y\V^0}_2 \geq \delta^0_{[s]}\}$, where $\wt{\x}_j$ is the $j$th column of $\X$ and $\delta_{[1]}^0 \geq \delta_{[2]}^0 \geq \dots \geq \delta_{[p]}^0$ is the descending sort of $\{\delta^0_j\}$ with $\delta^0_j=\norm{\wt{\x}_j\trans\Y\V^0}_2$. That is equivalent to update the active set by the formula $\{1\leq j \leq p: \delta_j^{0} \geq \delta_{[s]}^{0}\}$ with the given $\V^0$ and $\B^0=\0$. 

\begin{algorithm}[!htp]
    \caption{Multi-response Best Subset Selection with fixed $r$ and $s$ (MrBeSS.fixed)}\label{alg:mrbess}
    \hspace*{\algorithmicindent} \textbf{Input}: response matrix $\Y$, predictor matrix $\X$,  rank $r$  and row sparsity $s$.  
    \begin{algorithmic}[1]
    \State \textbf{Initialization}: set $k=0$ and initialize the active set $A^{k}$.
    \Repeat
    \State Perform eigen-decomposition to $\Y\trans\X_{\cdot A^k}(\X_{\cdot A^k}\trans\X_{\cdot A^k})^{-1}\X_{\cdot A^k}\trans\Y$, and let $\V^{k+1}$
    \hspace*{\algorithmicindent} $\in\mbR^{q\times r}$ consist of eigenvectors corresponding to the top-$r$ eigenvalues. 
    \State Update the primal and dual variables ($\B^{k+1},\bGamma^{k+1}$) by the formula \eqref{eq:kkt:approx}. 
    \State Calculate the sacrifice $\{\delta_j^{k+1} = \norm{\b_j^{k+1}+\bgamma_j^{k+1}}_2: 1\leq j\leq p\}$.
    \State Determine the active set $A^{k+1}$ by $\{1\leq j \leq p: \delta_j^{k+1} \geq \delta_{[s]}^{k+1}\}$.
    \State Update the coefficient matrix by $\C^{k+1} = \B^{k+1}(\V^{k+1})\trans$.
    \State Set $k \leftarrow k + 1$.
    \Until{$\norm{\C^{k+1} - \C^k}_{F}\leq \tau$, where $\tau$ is a user-specified tolerance, e.g. $\tau=10^{-5}$.}
    \end{algorithmic}
    \hspace*{\algorithmicindent} \textbf{Output}: $\wh{\C}=\C^k$. 
\end{algorithm}

The existing methods, e.g. \citet{she2017selective,bunea2012joint,chen2012sparse,uematsu2019sofar}, usually formulate the sparse reduced rank regression as a nonconvex optimization by ``loss function + regularization''. Then they use the block relaxation algorithm to solve the nonconvex optimization that only attains a local optimum and is difficult to ensure the statistical convergence of the numerical output. Different from these approaches, MrBeSS iteratively updates the primal variable $\B\V\trans$ and the dual variable $\bGamma$ and then updates the support set by a hard thresholding rule, which is an ``implicit regularization''. Owing to this framework, it enables us to analyze the statistical properties in the view of the algorithm. Recently, we notice that there have be some literature to analyze the statistical convergence in the view of the algorithm, e.g. \citet{qiu2022} and \citet{zhao2022high}. But MrBeSS is the first to solve the sparse reduced rank regression. In addition, different from the gradient descent method \citep{qiu2022,zhao2022high}, MrBeSS needs not to choose the step size for the algorithm. In the following theoretical analysis, we reveals that the numerical solution of MrBeSS enjoys the nearly optimal convergence rate, which bridges the gap between the statistical guarantee and numerical computation. Moreover, we will show that the iteration complexity is polynomial to attain the nearly optimal estimation. As far as we know, this is the first to the sparse reduced rank regression. 




\subsection{MrBeSS with adaptive parameter tuning}
In practice, the rank $r$ and the row-sparsity $s$ are unknown, and some data-driven procedure is needed to tune them. One approach is to select parameters by the $k$-fold cross validation technique \citep{hastie2009elements} based on a two-dimensional grid of $s$ and $r$. Nevertheless, it is time-consuming especially with a large grid. Alternatively, 
we propose a generalized-type information criterion (GIC) to tune parameters. For any coefficient matrix $\C$,  we define GIC as follows:
\begin{align}
    \text{GIC}(\C) = (2n)^{-1}\norm{\Y - \X\C}_F^2 + q(s+r)(\log\log n)\{n^{-1}\log p\}^{1/2}, 
    \label{def:GIC}
\end{align}
where $s$ is the number of non-zero rows of $\C$ and $\rank{\C}=r$. The GIC attempts to trade off between the prediction error and the complexity of the model, and smaller GIC represents a more balanced choice. 

A naive approach is to minimize GIC among all candidate parameters over a grid of $r$ and $s$. However, simultaneous search for optimal pair of tuning parameters over a two-dimensional grid is still computationally expensive. For instance, if the numbers of the candidates for $s$ and $r$ are $20$ and $10$, then we need to calculate $100$ kinds of combination of $s$ and $r$. Therefore, to reduce the computational burden, we introduce a simplified yet efficient search strategy. In specific, we first identify a GIC-minimal choice of sparsity $\wh{s}$ over a sequence of $s$ values with rank $r_{\max}$ large enough, and then select an optimal rank with the minimum GIC value by fixed $s=\wh{s}$. The algorithm with the simplified search strategy is summarized in \textbf{Algorithm \ref{alg:amrbess}}, which is shown by Theorem~\ref{theorem:3} to recover the true tuning parameters.

\begin{algorithm}[!ht]
    \caption{Multi-response Best Subset Selection (MrBeSS)}\label{alg:amrbess}
    \hspace*{\algorithmicindent}  \textbf{Input}: response matrix $\Y$, predictor matrix $\X$, maximum number of rank $r_{\text{max}}$\\
    \hspace*{\algorithmicindent} \hspace*{\algorithmicindent} \hspace*{\algorithmicindent} and row sparsity $s_{\text{max}}$.  
    \begin{algorithmic}[1]
    \For{$s= 1,\dots, s_{\text{max}}$}
        \State {Run \textbf{Algorithm 1} with $r_{\max}$ and $s$. Denote the output by $\wh{\C}^{s, r_{\max}}$.}
        \State Compute the GIC value $\text{GIC}(\wh{\C}^{s, r_{\max}})$ via \eqref{def:GIC}.
      \EndFor
    \State Determine the optimal sparsity by $\wh{s} = \arg\min_{s} \text{GIC}(\wh{\C}^{s}).$
    \For{$r = 1, \dots, r_{\max}$}
        \State {Run \textbf{Algorithm 1} with $r$ and $\wh{s}$. Denote the output by $\wh{\C}^{\wh{s}, r}$.}
        \State Compute the GIC value $\text{GIC}(\wh{\C}^{\wh{s}, r})$ via \eqref{def:GIC}.
      \EndFor
    \State Determine the optimal rank by $\wh{r} = \arg\min_{r} \text{GIC}(\wh{\C}^{\wh{s}, r}).$
    \State Run \textbf{Algorithm 1} with $\wh{r}$ and $\wh{s}$. Denote the output by $\wh{\C}$.
    \end{algorithmic}
    \hspace*{\algorithmicindent} \textbf{Output}: $\wh{\C}$. 
  \end{algorithm}
  
%
%

%% file: theory.tex
\section{Theoretical justification}\label{sec:theory}

In this section, we present the theoretical analysis for the estimators generated by \textbf{Algorithm~\ref{alg:mrbess}} and \textbf{Algorithm~\ref{alg:amrbess}}. The proof of these results are provided in Appendix.

\subsection{Technical conditions}

First, we introduce some definitions that will be used later. For any output $\wh{\C}$ of \textbf{Algorithm~\ref{alg:amrbess}}, denote the identified support set for rows as $\wh{A} = \left\{1 \leq i \leq p: \norm{\wh{\c}_{i}}_2 \neq 0\right\}$, where $\wh{\c}_{i}$ is the $i$th row of $\wh{\C}$. Denote the rank of $\wh{\C}$ as $\wh{r} = \rank{\wh{\C}}$. Throughout this section, we search the solution in the following parameter space for the coefficient matrices
\begin{align}
    \mathcal{C}(K) = \left\{\C\in\mbR^{p\times q}:~\norm{\C}_{2,0}\leq s,~\rank{\C}\leq r,~\text{and}~r\leq s \leq K\right\},
    \label{def:para-space}
\end{align}
where $K\geq s^* \geq r^*$ and can be diverged with the sample size $n$. We assume the following technical conditions for the theoretical analysis.

\begin{condition}\label{cond:1}
    With the given sparsity level $s$, there is
    \begin{align*}
        0 < c_{-}(s) \leq \inf_{\u\neq \0} \frac{\norm{\X_{\cdot A} \u}_2^2}{n\norm{\u}_2^2}
        \leq \sup_{\u\neq \0} \frac{\norm{\X_{\cdot A} \u}_2^2}{n\norm{\u}_2^2} \leq c_{+}(s) < +\infty,
    \end{align*}
    with an arbitrary subset $A\subset \{1,2\dots,p\}$ and $\abs{A} \leq s$.
\end{condition}

\begin{condition}\label{cond:2}
    With the given sparsity levels $s$, there is
    \begin{align*}
        \sup_{\u\neq \0} \frac{\norm{\X_{\cdot B}\trans\X_{\cdot A} \u}_2}{n\norm{\u}_2} \leq \theta_{s},
    \end{align*}
    with two arbitrary subsets $A,B\subset\{1,2\dots,p\}$, $\abs{A}\leq s$, $\abs{B}\leq s$ and $A\cap B = \varnothing$.
\end{condition}

\begin{condition}\label{cond:3}
    Assume that the rows of $\E$ are i.i.d from the normal distribution $N(\0,\bSigma)$, and the operator norm of $\bSigma$ is bounded from the above.
\end{condition}

\begin{condition}\label{cond:4}
    With the $s\geq s^*$ and $r > r^*$, define the parameter $\gamma$ as follows
    \begin{align*}
        \gamma
        = \frac{\theta_{s}(1+\sqrt{r})(1+\theta_{s})}{c_{-}^2(s)} + \frac{(1+\sqrt{r})\theta_{s}}{c_{-}(s)},
    \end{align*}
    and assume that $\gamma < 1$.
\end{condition}

Condition \ref{cond:1} is the restricted eigenvalue (RE), e.g. \citet{zheng2019scalable,bickel2009simultaneous}, where the upper bound $c_{+}(s)$ usually holds with the normalized data, and $s$ implies the value $c_{+}(s)$ will slightly increase with the cardinality $s$. The cardinality $s$ is small in practice thus the upper bound will be finite. As for the lower bound $c_{-}(s)$, it ensures the identifiability of the model. $c_{-}(s) > 0$ is equivalent to the matrix $\X_{\cdot A}\trans\X_{\cdot A}$ is inversible. Once $A$ is the true support set, we can directly implement the ordinary reduced rank estimation, which can be seen as the oracle estimation. Later, we will reveal that the numerical solution from the proposed algorithm is equal to the oracle estimation with a significant probability. 

Condition \ref{cond:2} is a derivative of the upper bound in Condition \ref{cond:1}. We define the notation for the algorithm analysis. In fact, there is $\theta_s\leq c_{+}(s)$. Compared with Condition \ref{cond:1}, the difference is that $\theta_s$ describes the correlation among of different predictors. The larger $\theta_s$ results in a more challenging recovery for the coefficient, since the strong pseudo-correlation misleads the variable selection. A special case is $\theta_s = 0$, which means the columns of $\X$ are mutually orthogonal. Under the case, there is a closed formulation for the optimization with nonconvex regularization that is easy to solve. In this paper, we concerns on the case $\theta_s > 0$. 

Condition \ref{cond:3} is a regular condition to analyze the statistical convergence rate, where we use the Gaussian probability tail to bound the magnitude of noise. Compared with the previous work such as \citet{bunea2012joint,she2017selective}, we allow the correlation among of the entries in the noise vector. This is more reasonable since the elements in the response vector are correlated with each other in practical applications.

We define a parameter $\gamma$ in Condition \ref{cond:4} and impose that $\gamma < 1$ with $s \geq s^*$. \citet{hunag2017L0} has shown that $\theta_s \leq \big(c_{+}(s) - 1)\vee (1-c_{-}(s)\big)$. Then with some appropriate $c_{-}(s)$ and $c_{+}(s)$ (e.g. $c_{-}(s)\geq 0.87$ and $c_{+}(s)\leq 1.16$ with $r=3$), we have $\gamma < 1$. There is a similar parameter to $\gamma$ in the univariate-response regression, where the only difference is the rank factor $\sqrt{r}$ replaced by an absolute constant \citep{hunag2017L0,zhu2020polynomial}. In a simple point of view, the $\gamma$ will be smaller than 1 as long as the correlation among of the different covariates is small enough, in other words, $\theta_{s}$ is as small as enough. The difference between the univariate response and the multiple lies in the rank structure, where we need a smaller correlation with a higher rank. The parameter $\gamma$ can be seen as a generalized version for the sparse reduced-rank regression. 

\begin{definition}\label{def:2}
    Define the signal strength of $\C^*$ as follows
    \begin{align*}
        \delta_n = \big\{\underset{A^*\not\subset A, |A| \leq K }{\inf}~n^{-1}\norm{(\I - \bP_A)\X\C^*}_F^2\big\} \wedge \sigma_{r^*}^*,
    \end{align*}
    where $\sigma_{r^*}^*$ is the $r^*$th largest eigenvalue of $n^{-1}\C\strans\X\trans\X\C^*$, and $\bP_A$ is the projection matrix onto the space spanned by the columns of $\X_{\cdot A}$.
\end{definition}

In Definition \ref{def:2}, the first term in $\delta_n$ measures the smallest signal strength of the true variables, which is a matrix version of the definition in \citet{fan2013tuning} for the recovery of the support set $A^*$. The second term in $\delta_n$ describes the signal strength on the singular value for the recovery of the true rank, and similar definitions have been proposed for the recovery of the rank in reduced-rank regression models, see \citet{zheng2019scalable,bunea2011optimal}. 

\subsection{Main results}

Now we are ready to present the main result. The first theorem guarantees the convergence of \textbf{Algorithm \ref{alg:mrbess}}, where the detail is as follows.  

\begin{theorem}\label{theorem:1}
    Denote $\C^{k+1}$ as the $k$th iteration of \textbf{Algorithm \ref{alg:mrbess}} with $s \geq s^*$ and $r\geq r^*$. Assume that Conditions \ref{cond:1}--\ref{cond:4} hold, then the following inequalities
    \begin{align*}
         & \norm{\C^{k+1} - \C^*}_{\op}
        \leq \gamma^k \left(1 + \frac{\theta_{s}}{c_{-}(s)}\right) \norm{\C^*}_{\op} + O\left(n^{-1/2}\{s+q+s\log(ep/s)\}^{1/2}\right),
        \\
         & \norm{\C^{k+1} - \C^*}_{F}
        \leq \sqrt{2r} \gamma^k \left(1 + \frac{\theta_{s}}{c_{-}(s)}\right) \norm{\C^*}_{\op} + O\left(n^{-1/2}\{r(s+q)+rs\log(ep/s)\}^{1/2}\right),
    \end{align*}
    hold with the probability at least $1-p^{-c}$, where $c$ is a positive constant.
\end{theorem}

Theorem \ref{theorem:1} shows that the solution from \textbf{Algorithm \ref{alg:mrbess}} converges geometrically to the truth $\C^*$ with $\gamma < 1$. Moreover, with the large enough $k$, we can obtain the statistical convergence rate, which is presented in the following corollary. 

\begin{corollary}\label{corollary:1}
    With the same conditions in Theorem \ref{theorem:1} and $\omega=\left(1 + \frac{\theta_{s}}{c_{-}(s)}\right)\norm{\C^*}_{\op}$, the below oracle inequalities 
    \begin{align*}
        &\norm{\C^{k+1}-\C^*}_{\op} \leq O\left(n^{-1/2}\{s+q+s\log(ep/s)\}^{1/2}\right), 
        \\
        &\norm{\C^{k+1} - \C^*}_F \leq O\left(n^{-1/2}\{r(s+q)+rs\log(ep/s)\}^{1/2}\right),
    \end{align*}
    hold uniformly with the probability at least $1-p^{-c}$ when the number of iterations satisfy 
    \begin{align*}
        k = O\left(\log_{\gamma^{-1}}\left\{\frac{n\omega^2}{s+q+s\log(ep/s)}\right\}\right)\,\,\text{with some}\,\, 0 < \gamma < 1. 
    \end{align*}
\end{corollary}

First, we emphasize that $\C^{k+1}$ is the numerical output of the algorithm that is different from the current literature, where they concern on the statistical property of the global optimum of an optimization. However, the nonconvexity of the sparse reduced rank regression results in their numerical algorithms only attaining a local optimum. There exsits a gap between the computational method and the theoretical guarantee in the current literature. Recall the above corollary, the left of the inequality is the numerical error of the algorithm, and the right is the statistical convergence rate. Therefore, the corollary bridges the gap between the numerical computation and the theoretical analysis. 

Second, compared with the minimax lower bound proposed by \citet{ma2020adaptive,ma2014adaptive}, the statistical convergence rate of $\C^{k+1}$ has been nearly minimax optimal. Although there have been some methods attaining the optimal rate \citep{she2017selective,uematsu2019sofar}, their numerical results usually are some local optimums that leads to a gap between numerical solutions and their optimal convergence rate. As far as we know, the proposed algorithm in this paper is the first, where its numerical output enjoys the nearly minimax optimal rate. 

Third, owing to the bridge of the above corollary, we can obtain the iteration complexity of the proposed algorithm, which is also a new result compared with the current methods. The corollary reveals that the proposed algorithm only needs a polynomial iteration complexity to attain the optimal estimation. Thereby, it enjoys a fast computation in high dimensions. 

Last but not least, we establish the error bound with respect to the operator norm that is commonly used in the matrix perturbation. It enables us to obtain a tighter upper bound for the matrix decomposition than the Frobenius norm. This result also distinguishes our work from the current literature.

Theorem \ref{theorem:1} and Corollary \ref{corollary:1} are built on the output of \textbf{Algorithm} \ref{alg:mrbess}, say $\C^{k+1}$, where the algorithm depends on two hyper-parameters, $r$ and $s$, which need be tuned. Once given with the ture $r^*$ and $s^*$, we can obtain a very accurate estimation, together with the optimal convergence rate. Therefore, we propose a new information criterion for the tuning of $s$ and $r$, and then it yields an oracle estimation stated in Theorem \ref{theorem:3}. We first introduce the property of the proposed GIC with regrad to the numerical solution in \textbf{Algorithm} \ref{alg:mrbess}. Note that $\C^{k+1}$ actually is the reduced-rank estimator \citep{Reinsel1998} restricted in the corresponding support set. Thus we consider an arbitrary reduced-rank estimator $\wh{\C}$ in Theorem \ref{theorem:2}. Let the oracle reduced-rank estimator $\wh{\C}^{\ora}$ be the baseline. The theorem reveals that the GIC arrives at the minimum only when $\wh{\C} = \wh{\C}^{\ora}$ with the large enough $n$.

\begin{theorem}\label{theorem:2}
    Under Conditions \ref{cond:1}--\ref{cond:4}, assume that the following assumptions hold,
    \begin{align*}
        &K\{n^{-1}(q + K\log p)\}^{1/2}\norm{\C^*}_{\op} = o(\delta_n), 
        \\ 
        &K q(\log\log n)\{n^{-1}\log p\}^{1/2} = o(\delta_n),
        \\
        & K(\log p)^{-1/2}\norm{\C^*}_{\op} = o(q^{1/2}\log\log n).
    \end{align*}
    Then with the large enough $n$ and arbitrary reduced-rank estimator $\wh{\C}\in\mathcal{C}(K)$ with the support set $A$ and $\rank{\wh{\C}}=r$, we have
    \begin{align*}
        \bP\left\{
        \underset{A\neq A^*~\text{or}~r\neq r^*}{\inf}\GIC(\wh{\C}) > \GIC(\wh{\C}^{\ora})
        \right\}
        \geq 1 - p^{-c},
    \end{align*}
    where $\wh{\C}^{\ora}$ is the oracle reduced-rank estimator with the support set $A^*$ and $\rank{\wh{\C}^{\ora}} = r^*$.
\end{theorem}

The above theorem implies we can obtain the oracle estimator by the proposed GIC asymptotically. In the recent literature, \citet{she2017selective} also proposed a predictive information criterion (PIC) to tune the rank and the support set but there is fundamentally different from our GIC. In this paper, the proposed GIC aims to the consistency of the recovery for the rank and the true support set, however, PIC aims to obtain an accurate prediction error and it cannot ensure the consistency of the estimation for the rank and support set. 
Owing to the GIC, we can show that the numerical output of algorithm actually is the oracle estimation with a significant probability, which is presented in the next theorem. 

Although Theorem \ref{theorem:2} reveals that GIC can asymptotically recover the support set and rank, we still need to search parameters over the grid generated by $s$ and $r$. To reduced the searching range, we develop \textbf{Algorithm \ref{alg:amrbess}} to tune parameters along with the coordinate. That is, we first derive an optimal sparsity level $\wh{s}$ under a given relatively large rank, say $r_{\max}$, and then tune the rank by GIC with the given $\wh{s}$. The following theorem provides the theoretical justification for this strategy.

\begin{theorem}\label{theorem:3}
    Denote $\wh{\C}$ as the output of \textbf{Algorithm \ref{alg:amrbess}} with $s^*\leq s_{\max}$, $r^* \leq r_{\max}$ and $\max\,\{s_{\max},r_{\max}\}\leq K$. Under the same conditions of Theorem~\ref{theorem:2} and sufficiently large $n$, there is 
    \begin{align*}
        \bP\left\{\wh{\C} = \wh{\C}^{\ora}\right\} \geq 1- p ^{-c}.
    \end{align*}
\end{theorem}

Theorem \ref{theorem:3} reveals an interesting result that the output of \textbf{Algorithm \ref{alg:amrbess}}  is exactly the oracle estimator with an overwhelming probability. Different from the existing literature, Theorem \ref{theorem:3} guarantees the statistical property of the algorithmic solution. In the sparse reduced rank regression, the coordinate descending algorithm commonly used in the current literature \citep{she2017selective,ma2014adaptive,mishra2017sequential} converges to a local minimizer but there is no literature to discuss the relationship between the numerical solution and the oracle reduced rank regression. The oracle estimation is an important baseline since it is an ideal result for the variable selection and the dimension reduction. The result implies that the obtained estimation is unbiased with an overwhelming probability. The numerical results in Section~\ref{sec:simulation} also show that the estimation error of MrBeSS is lower than those of other competing methods. 


%% file: simulation.tex
\section{Numerical Studies}\label{sec:simulation}

{
In this section, we investigate the finite-sample performance of MrBeSS on simulated data and compare it with other competing methods including the reduced rank regression via adaptive nuclear norm penalization (RRR-ada, \citet{chen2013reduced}), the rank constrained group lasso (RCGL, \citet{bunea2012joint}), the sparse reduced rank regression using adaptive group lasso (SRRR, \citet{chen2012sparse}), and the sequential factor extraction via co-sparse unit-rank estimation (SeCURE, \citet{mishra2017sequential}). 
}

\subsection{General setups}\label{sec:simulation:setting}

We first introduce some general notations and simulation settings, which will be used later. 
To generate the coefficient matrix, we first generate matrix $\A\in\mbR^{q\times r^*}$, whose entries of matrix  are i.i.d. from $N(0,1)$. Then, let the entries of matrix $\B_1\in\mbR^{s\times r^*}$ be i.i.d. from the uniform distribution on $[-1,-0.3]\cup [0.3,1]$, and we can construct matrix $\B$ as $\B = [\B_1\trans,\0_{(p-s)\times r^*}\trans]\trans\in\mbR^{p\times r^*}$. {We normalize the columns of $\A$ and $\B$ such that the $\ell_2$ norm of each column is unit-length, respectively. Moreover, we generate an $r^*\times r^*$ diagonal matrix whose the $k$th diagonal entry $d_k = d_0 + 5k$ with $d_0=5$.  After that, we set the coefficient matrix as 
\begin{align*}
    \C^* = \B\D\A\trans \in \mbR^{p\times q},
\end{align*}
where $\D=\diag{d_1^*,\cdots,d_{r^*}^*}$ is the $r^*\times r^*$ diagonal matrix. It is easy to see that the rank of $\C^*$ is equal to $r^*$.}

We generate the predictor matrix $\X=\left[\x_1,\dots,\x_n\right]\in\mbR^{n\times p}$ from the normal distribution. In particular, the $i$th row $\x_i$ is independently sampled from $N(\0,\bGamma)$ with $\bGamma = (\gamma_{ij})\in\mbR^{p\times p}$ and $\gamma_{ij}=0.5^{\abs{i-j}}$. Similarly, the rows of the noise matrix $\E$ are generated as i.i.d. samples of $N(\0,\omega\bSigma)$ with $\bSigma=(\sigma_{ij})\in\mbR^{q\times q}$ and $\sigma_{ij}=0.3^{\abs{i-j}}$. We set $\omega$ to control the signal-to-noise ratio (SNR), defined as $\text{SNR} = \eta_{r^*}/\norm{\E}_F$, where $\eta_{r^*}$ is the $r^*$ largest eigenvalue of $\X\C^*$. Moreover, in this paper, we consider two kinds of $\bSigma$ for the noise matrix: (1) the auto regression matrix (AR) satisfying $\sigma_{ij} = 0.3^{\abs{i-j}}$; (2) the strong correlation matrix (SC) satisfying $\sigma_{ij}=0.3$ if $i\neq j$, otherwise, $\sigma_{ij}=1$. With the given $\X$, $\C^*$ and $\E$, the response matrix is $\Y = \X\C^* + \E$. 
We set $n=100, q=100, s=10, r^*=3, \text{SNR}=0.5$, and let $p$ be varied from 200 to 1000. For each setup, a total of 200 replications are conducted. 

{
For any estimated coefficient matrix $\widehat{\C}$, we measure the estimation and prediction accuracy by  $\text{Er}(\widehat{\C}) = \norm{\widehat{\C}-\C^*}_{F}^2/(pq)$, and $ \text{Er}(\X\widehat{\C}) = \norm{\X(\widehat{\C} - \C^*)}_{F}^2/(nq)
$, respectively. The variable selection performance is characterized by the false positive rate (FPR) and false negative rate (FNR) in recovering the sparsity patterns of the row support set of $\C^*$, where $\text{FPR} = \text{FP}/(\text{TN}+\text{FP})$ and $\text{FNR} = \text{FN}/(\text{TP}+\text{FN})$ with TP, FP, TN and FN being the numbers of true nonzeros, false nonzeros, true zeros, and false zeros of the rows of $\C^*$, respectively. In addition, we also report the estimated rank and the computational time (Time) in seconds.

To tune parameter(s), we search over a grid of sparsity and rank for all methods except RRR-ada, and a line search of rank for RRR-ada. We set the maximum rank as 10 when tuning parameters for all methods.
Because there is no explicit information criteria for RCGL and SRRR under high dimensions, we tune the parameters by data validation for them. That is, we use $80\%$ samples to estimate coefficients and the other samples to measure the prediction with different combinations of sparsity and rank. Then we can determine the optimal pair of sparsity and rank with the minimum prediction error. 
As for RRR-ada and SeCURE, we follow the default criteria to tune the parameter, which can be found in \citet{chen2013reduced,mishra2017sequential} for more details.
For a fair comparison, we present two versions of our proposal denoted by MrBeSS-V and MrBeSS. While in MrBeSS-V the optimal parameters are determined by data validation, MrBeSS tunes parameters based on GIC. 
All of the competing methods except SeCURE are implemented in the R package \code{rrpack}, and SeCURE are implemented in R package \code{secure}. 
}

\subsection{Simulation results}

The simulation results are summarized in Table~\ref{tab:1}. It can be seen that all methods except RRR-ada and SeCURE have comparable performance in terms of estimation and prediction errors. This is due to RRR-ada doesn't perform variable selection at all, and thus it leads to inaccurate estimation for the coefficient matrix in the sparse data setting. As for SeCURE, it imposes the sparsity on both rows and columns, and the performance is not well for the row-sparse setting. This can be seen from the high values in FNR and biased estimation in rank, which indicates it fails to recover the true nonzero rows nor the hidden structure in the coefficient matrix.

Among the methods suitable for row-sparse scenarios, both MrBeSS and MrBeSS-V have substantial better performance. In particular, MrBeSS-V yields the sparsest model with much lower values in FPR and comparable FNR values in comparison with SRRR and RCGL, the methods also using validation data to tune parameters. In other words, both SRRR and RCGL have the tendency of over-selecting irrelevant variables as indicated by the high FPR values, which is expected since a Lasso-type penalty is used to perform row selection. This shows the superiority of restricting the number of nonzero rows for variable selection in the reduced rank regression model. In addition, the computational time of MrBeSS-V is less than other two methods, and the gap is growing as $p$ increases. 

From Table~\ref{tab:1}, we can tell that both MrBeSS and MrBeSS-V have comparable performance when $p$ is low. When $p$ is high, say $p=1000$, MrBeSS-V slightly outperform MrBeSS in terms of FNR, which might be due to the small sample size ($n=100$) compared with the dimension. Nevertheless, the computational time for MrBeSS is less than one tenth of those for MrBeSS-V, showing its feasibility in dealing with high-dimensional data.

Overall, the two MrBeSS approaches produce considerably sparser model with adequate estimation and prediction accuracy among all methods, and enjoy nice computational efficiency especially in high dimensions.

\begin{table}
    \caption{The average results with AR covariance matrix in $\E$, with their standard errors in parentheses. \label{tab:1}}
    \centering 
    \resizebox{0.9\textwidth}{!}{
        \input{tables/table-comparision.tex} 
    }
\end{table}

\subsection{Tuning strategy in MrBeSS}

In this section, we show the effectiveness of using GIC to tune parameters by comparing two versions of our approach, i.e., MrBeSS and MrBeSS-V, under different setting of sample size $n$. In particular, we adopt settings from Section~\ref{sec:simulation:setting} and vary the sample size $n$ from 100 to 400 with the step size 100. For space limit, we only present the results of $p=400, 600, 800$ and the noise with the SC covariance matrix, and the remaining results to be omitted since they are similar.

\begin{figure}[!htp]
    \centering
    \includegraphics[width=0.9\textwidth]{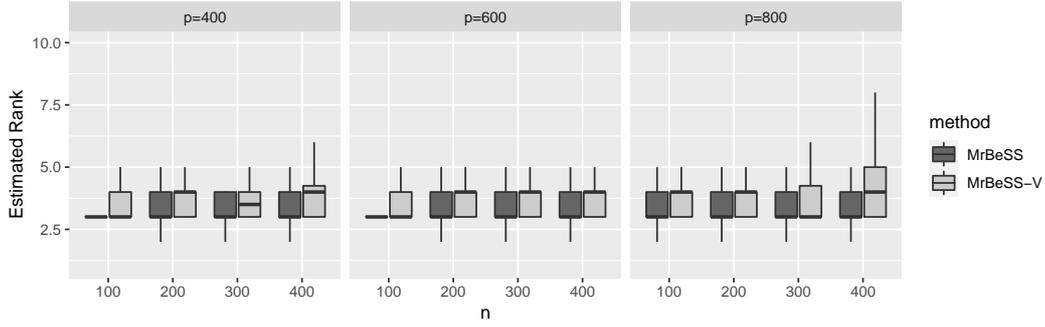}
    \caption{The estimated rank by MrBeSS and MrBeSS-V. \label{fig:rank}}
\end{figure}

First the result of the rank estimation is presented in Figure \ref{fig:rank}.
From Figure \ref{fig:rank}, we can see that the proposed GIC is conservative for the rank estimation under the low sample size but it can estimate the rank accurately when increasing the sample size. The result coincides with the theory in Theorem \ref{theorem:3} that the GIC can recover the rank asymptotically with the large enough $n$. 

\begin{figure}[!htp]
    \centering
    \includegraphics[width=0.9\textwidth]{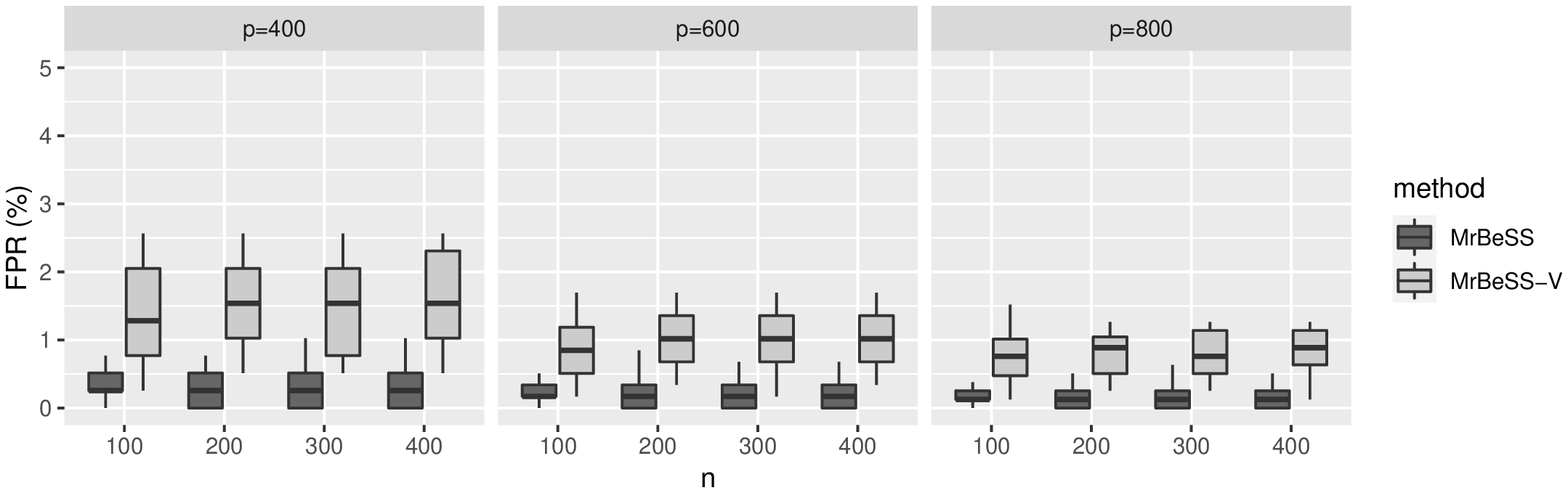}
    \includegraphics[width=0.9\textwidth]{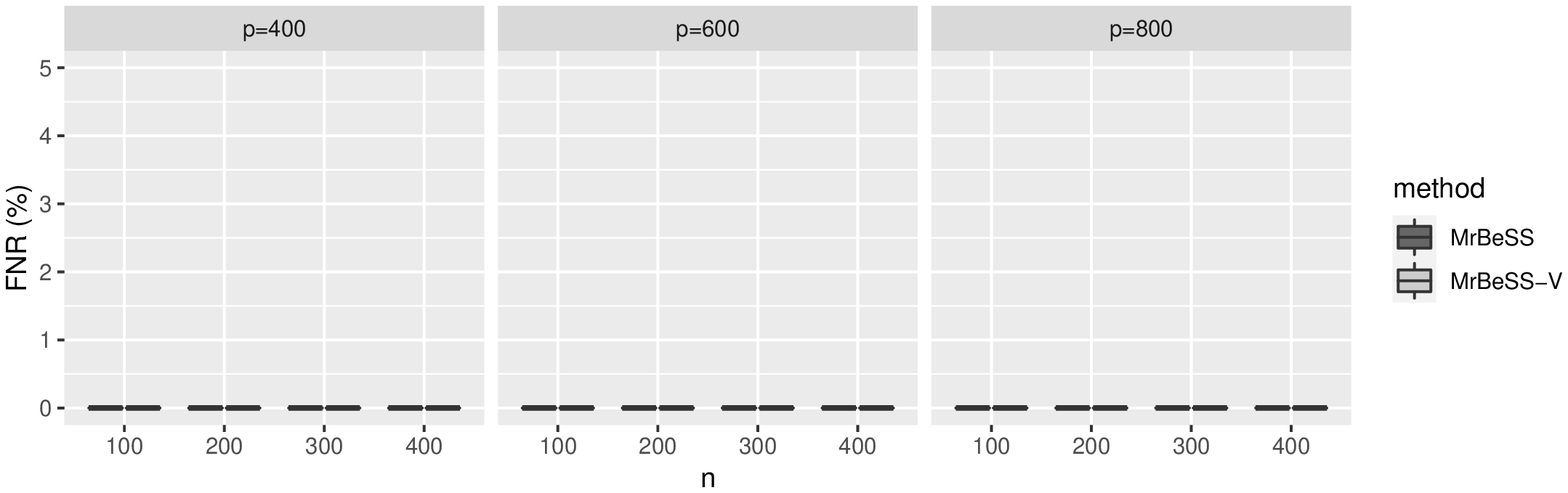}
    \caption{The performance of MrBeSS and MrBeSS-V in the variable selection.\label{fig:var-selection}}
\end{figure}

Next the variable selection performance is attached in Figure \ref{fig:var-selection} that shows the GIC is outstanding in the variable selection. Even under the low sample size, GIC still recovery the support set accurately. Compared with the GIC, the method of data validation is conservative in the variable selection because it prefers to the solution that enjoys accurate prediction. This is a common issue for the data validation framework. 

\begin{figure}[!htp]
    \centering 
    \includegraphics[width=0.9\textwidth]{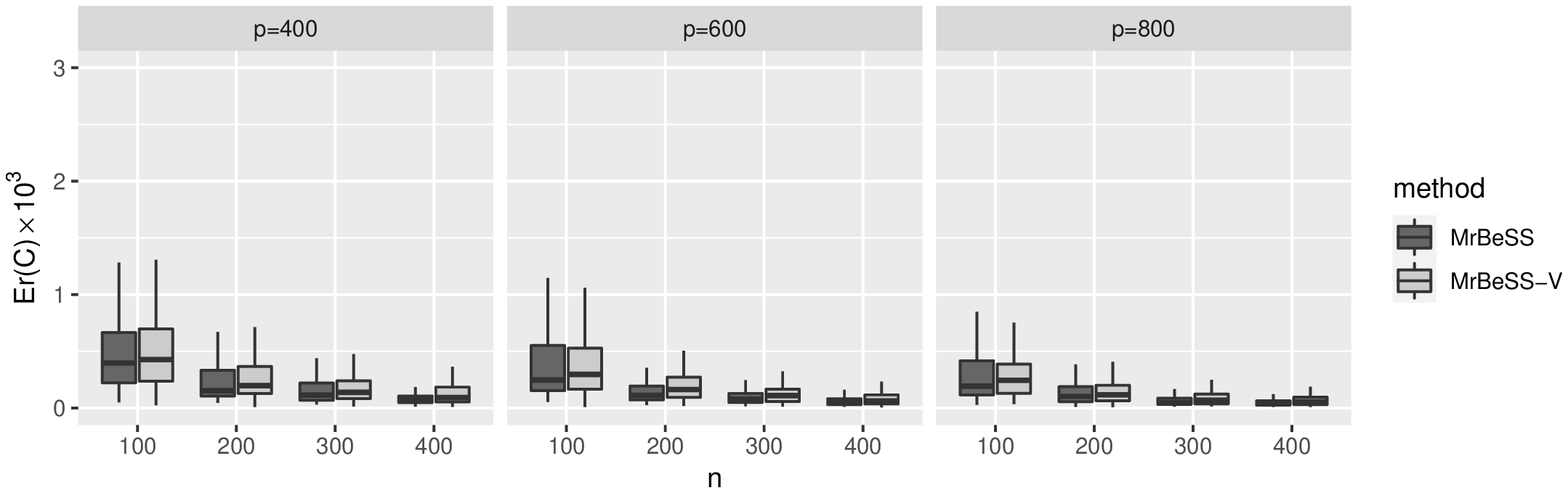}
    \includegraphics[width=0.9\textwidth]{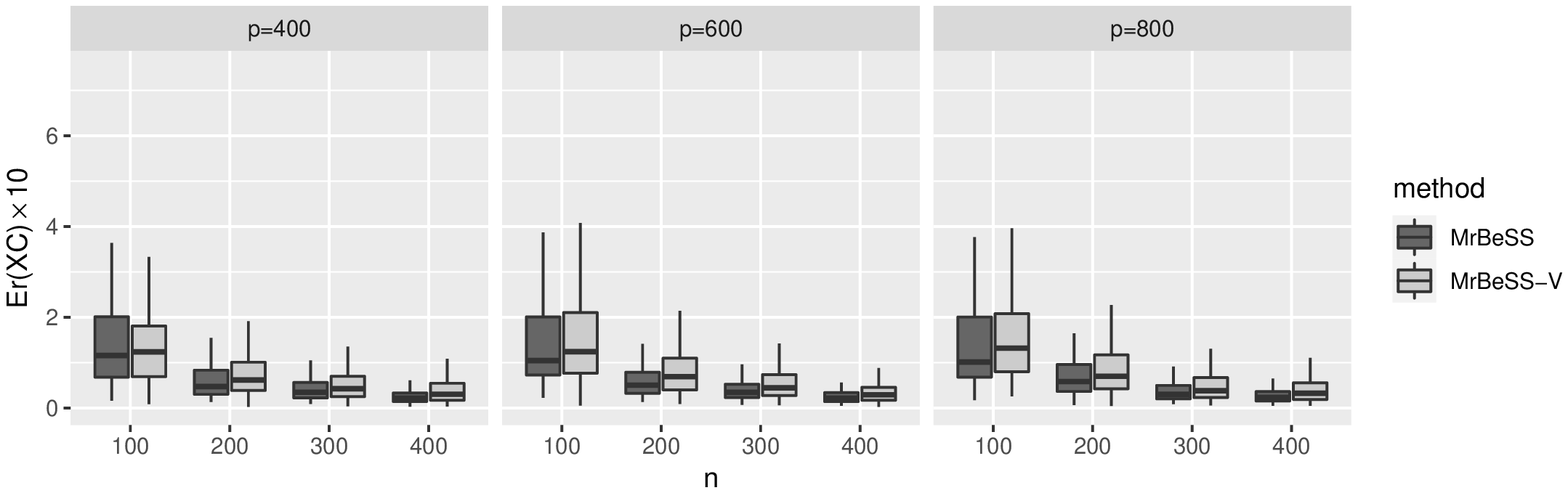}
    \caption{The estimation and prediction performance of MrBeSS and MrBeSS-V.\label{fig:est-pred}}
\end{figure}

Finally, the performance of estimation and prediction is presented in Figure \ref{fig:est-pred}. With the increasing sample size $n$, the accuracy of estimation and prediction is enhanced, which demonstrates the asymptotic consistency for both two versions of MrBeSS.

%% file: tables/table-comparision.tex
\begin{tabular}{lccccccc} 
\toprule 
& Method & Er$(\C)\times 1000$ & Er$(\X\C)\times 10$ & FPR (\%) & FNR (\%) & Time (s) & Estimated rank \\ 
\midrule 
 & \multicolumn{7}{c}{$\text{SNR}=0.5$} \\ 
\multirow{6}{*}{$p=200$} & RRR-ada & 19.24 (2.21) & 1.99 (1.62) & 100 (0) & 0 (0) & 0.24 (0.03) & 2.87 (0.36) \\ 
 & SRRR & 1.01 (0.5) & 1.16 (0.52) & 49.63 (5.79) & 0 (0) & 80.47 (5.11) & 3 (0) \\ 
 & RCGL & 0.93 (0.42) & 1.2 (0.53) & 67.11 (5.41) & 0 (0) & 65.08 (3.96) & 3 (0) \\ 
 & SeCURE & 23.84 (3.34) & 35.08 (7.37) & 0.64 (2.16) & 26.35 (15.95) & 0.38 (0.12) & 3.19 (0.84) \\ 
 & MrBeSS-V & 0.81 (1.36) & 1.08 (1.43) & 2.38 (1.34) & 0.65 (2.85) & 58.45 (9.38) & 3.02 (0.16) \\ 
 & MrBeSS & 1.24 (1.62) & 1.49 (1.53) & 0.67 (0.38) & 1.15 (3.35) & 2.54 (1.63) & 3.03 (0.59) \\ 
\hline 
\multirow{6}{*}{$p=400$} & RRR-ada & 13.95 (1.12) & 5 (7.9) & 98.5 (12.19) & 1.5 (12.19) & 1.28 (0.05) & 2.52 (0.73) \\ 
 & SRRR & 0.42 (0.35) & 1.07 (0.55) & 23.56 (4.77) & 0.05 (0.71) & 227.35 (13.12) & 3.04 (0.5) \\ 
 & RCGL & 0.47 (0.23) & 1.24 (0.62) & 40.18 (3.68) & 0 (0) & 177.45 (10.25) & 3 (0) \\ 
 & SeCURE & 12.72 (1.74) & 38.6 (8.07) & 0.34 (1.49) & 32.25 (16.82) & 1.35 (0.47) & 3.02 (0.79) \\ 
 & MrBeSS-V & 0.47 (0.84) & 1.2 (1.8) & 1.13 (0.68) & 1.05 (3.8) & 82.96 (13.25) & 3 (0.07) \\ 
 & MrBeSS & 0.71 (0.88) & 1.65 (1.7) & 0.35 (0.22) & 1.5 (3.98) & 3.92 (1.96) & 2.94 (0.5) \\ 
\hline 
\multirow{6}{*}{$p=600$} & RRR-ada & 10.24 (0.63) & 5.92 (8.64) & 98 (14.04) & 2 (14.04) & 3.86 (0.11) & 2.46 (0.78) \\ 
 & SRRR & 0.32 (0.46) & 1 (0.43) & 13.28 (2.91) & 0.35 (1.84) & 397.83 (23.23) & 3 (0) \\ 
 & RCGL & 0.35 (0.18) & 1.29 (0.54) & 29.34 (2.05) & 0 (0) & 301.62 (17.8) & 3 (0) \\ 
 & SeCURE & 8.52 (1.21) & 38.64 (7.81) & 0.08 (0.35) & 32.6 (17.43) & 3.28 (1.19) & 2.86 (0.53) \\ 
 & MrBeSS-V & 0.36 (1.12) & 1.51 (4.67) & 0.86 (0.45) & 1.2 (6.99) & 97.36 (14.66) & 3 (0) \\ 
 & MrBeSS & 0.52 (0.91) & 1.84 (2.73) & 0.22 (0.15) & 1.55 (5.85) & 4.88 (2.6) & 3.27 (1.15) \\ 
\hline 
\multirow{6}{*}{$p=800$} & RRR-ada & 8.1 (0.42) & 8.3 (12.33) & 96 (19.65) & 4 (19.65) & 8.89 (0.46) & 2.29 (0.91) \\ 
 & SRRR & 0.24 (0.34) & 0.95 (0.44) & 8.21 (2.2) & 0.35 (1.84) & 587.2 (32.98) & 3 (0) \\ 
 & RCGL & 0.28 (0.14) & 1.32 (0.61) & 22.96 (1.51) & 0 (0) & 442.25 (23.89) & 3 (0) \\ 
 & SeCURE & 6.54 (0.82) & 39.1 (7.66) & 0.07 (0.32) & 35.2 (16.32) & 7.05 (2.82) & 2.75 (0.55) \\ 
 & MrBeSS-V & 0.33 (0.94) & 1.81 (5.47) & 0.62 (0.33) & 1.75 (7.98) & 109.33 (17.75) & 3 (0) \\ 
 & MrBeSS & 0.43 (1.05) & 1.98 (4.21) & 0.19 (0.16) & 2 (9.08) & 6.05 (3.45) & 3.22 (1.13) \\ 
\hline 
\multirow{6}{*}{$p=1000$} & RRR-ada & 6.66 (0.38) & 11.62 (18.43) & 90 (30.08) & 10 (30.08) & 19.94 (5.82) & 2.15 (1.04) \\ 
 & SRRR & 0.24 (0.39) & 0.98 (0.51) & 6.02 (3.04) & 0.7 (2.75) & 782.44 (41.02) & 3.04 (0.51) \\ 
 & RCGL & 0.24 (0.11) & 1.33 (0.62) & 18.85 (1.16) & 0 (0) & 590.7 (32.43) & 3 (0.07) \\ 
 & SeCURE & 5.26 (0.64) & 39.62 (8.47) & 0.06 (0.22) & 33.9 (17.15) & 14.73 (7.88) & 2.71 (0.51) \\ 
 & MrBeSS-V & 0.25 (0.93) & 1.51 (4.72) & 0.53 (0.28) & 1.4 (8.68) & 120.43 (19.81) & 3 (0) \\ 
 & MrBeSS & 0.37 (0.86) & 2.21 (5.62) & 0.14 (0.12) & 2.3 (9.23) & 6.34 (3.06) & 3.13 (0.95) \\ 
\bottomrule 
\end{tabular} 

%% file: real-data.tex
\section{Real data}\label{sec:real}

We consider a gene expression and microRNA (miRNA) dataset from The Cancer Genome Atlas (TCGA) consortium \citep{cancer2011integrated}. We are interested in identifying miRNAs that regulate the expression of ovarian cancer related genes. Instead of using the whole set of 11,864 genes, we focus on a subset of 12 genes that have shown to be significantly associated with the four cancer subtypes \citep{cancer2011integrated}. The final dataset consists of $n=487$ samples with $q=12$ genes and $p=254$ measurements of miRNAs, after excluding the miRNAs with standard deviations less than 0.5.

We apply MrBeSS, as well as other four competing methods including RRR-ada, RCGL, SRRR, and SeCURE to these data. As for the tuning parameter, we use the same implementation in the previous section. Table~\ref{tab:real} reports the mean squared error $\text{MSE} = \frac1{nq}\|\Y - \X\wh{\C}\|_F^2$, the estimated rank and the estimated row-sparsity. We find from Table~\ref{tab:real} that while the classical method RRR-ada yields a null model, other methods result in a valid model with rank being non-zero. This suggests that RRR-ada might lose power in such high-dimensional data due to the horrible noise accumulation, which leads to RRR-ada mistakes the data all is noise. Both MrBeSS-V and SeCURE can achieve low prediction error, but the rank and the size of detected row support are much larger, which indicates a overfitting issue might occur in this specific data. This result is expect since MrBeSS-V is a validation method and would be conservative in the variable selection, see Figure~\ref{fig:est-pred}. Since the number of free parameters is determined jointly by rank and row-sparsity in RRR, MrBeSS successfully selects a model with simple structure as well as acceptable prediction accuracy. Furthermore, the running time of our MrBeSS is the lower among all methods, showing its practical application feasibility in high-dimensional data.

\begin{table}[!htp]
    \caption{Results of different methods on the ovarian cancer data.\label{tab:real}}
    \centering
    \resizebox{0.9\textwidth}{!}{
        \begin{tabular}{lcccc}
            \toprule
            Method   & Prediction Error & Time (s) & Estimated rank & Support set size \\
            \hline
            MrBeSS-V & 1.1              & 35.1     & 4              & 60               \\
            MrBeSS   & 1.55             & 0.51     & 1              & 10               \\
            SRRR     & 1.28             & 251.29   & 1              & 121              \\
            RCGL     & 1.23             & 236.37   & 1              & 235              \\
            RRR-ada  & 1.91             & 0.55     & 0              & 0                \\
            SeCURE   & 1.19             & 22.72    & 6              & 90               \\
            \bottomrule
        \end{tabular}
    }
\end{table}

In order to compare prediction accuracy and stability of different variable selection methods, we randomly split the data into a training set with $80\%$ samples and a test set with $20\%$ samples. All model estimations are carried out using the training data, with the parameter tuning strategies same as the previous experiment. We use the test data to calibrate the predictive performance of each estimator $\wh{\C}$, specifically, by its mean squared prediction error $\text{MSPE} = \frac1{n_{\text{test}}q}\|\Y_{\text{test}} - \X_{\text{test}}\wh{\C}\|_F^2$, where $\{\X_{\text{test}},\Y_{\text{test}}\}$ denotes the test data set. The random-splitting process is repeated 200 times with the results summarized in Table~\ref{tab:real:train}.
Again, MrBeSS-V and SeCURE have the smallest prediction error, yet yield the most complicate model. Besides these two methods, our MrBeSS approach has superior performance compared to the other methods in terms of prediction error and support set size. The SRRR and RCGL has competitive values in terms of prediction error, but their products of support set size and rank are around 13 times and 18 times of those in MrBeSS. In addition, the results for MrBeSS are most robust with the smallest standard errors in almost all measurements. We conclude that in real applications, MrBeSS can produce models with high prediction accuracy and interpretability.

\begin{table}[!htp]
    \caption{Results are based on splitting data into a training set with $80\%$ samples and a test set with $20\%$. Average results are reported over 200 replications, with their standard errors in parentheses.\label{tab:real:train}}
    \centering
    \resizebox{0.9\textwidth}{!}{
        \begin{tabular}{lcccc}
            \toprule
            Method   & Prediction Error & Time (s)       & Estimated rank & Support set size \\
            \hline
            MrBeSS-V & 1.47 (0.15)      & 40.29 (0.94)   & 4.76 (2.39)    & 44.6 (12.86)     \\
            MrBeSS   & 1.72 (0.16)      & 0.59 (0.14)    & 1.06 (0.25)    & 10.61 (0.85)     \\
            SRRR     & 1.81 (0.18)      & 173.98 (16.21) & 1.02 (0.19)    & 132.76 (27.3)    \\
            RCGL     & 1.85 (0.21)      & 210.44 (55.07) & 1.17 (0.49)    & 178.62 (90.59)   \\
            RRR-ada  & 1.91 (0.17)      & 0.62 (0.03)    & 0 (0)          & 0 (0)            \\
            SeCURE   & 1.51 (0.18)      & 16.46 (8.47)   & 3.72 (1.73)    & 59.06 (20.84)    \\
            \bottomrule
        \end{tabular}
    }
\end{table}

%% file: discussion.tex
\section{Discussion}\label{sec:dis}

In this paper, we propose a new algorithm, called MrBeSS, for the estimation of the sparse reduced-rank regression, where the approach mainly has the following advantages: (1) the method considers the $\ell_0$-norm constraint that avoids the bias in the estimation; (2) the algorithmic solution is nearly optimal with regrad to the statistical convergence rate; (3) the algorithm attains to the nearly optimal estimation with a polynomial number of iterations. Owing to the three advantages, MrBeSS enjoys not only nice sampling properties but also an accurate estimation and fast computation. Therefore, the performance of MrBeSS is better than other competitive methods.

There are some interesting problems for the future research. One extends the proposal to the case, where both rows and columns of coefficient matrix are sparse called the co-sparse structure \citep{mishra2017sequential}. Then we can implement the variable selection on responses and predictors simultaneously. Another one is to equip the primal-dual formulation with the unit-rank deflation, e.g. \citet{zheng2019scalable,chen2022}, that constructs the estimation for each rank-one component of the coefficient matrix. It can enhance the accuracy of estimation for each rank-one layer or singular vectors. Moreover, extending the algorithm to the multi-response generalized linear model is also important that has been widely used in recommender systems and reinforcement learning. 

%% file: appendix.tex
\appendix 

\section{Proofs of main results}

Before proving the main result, we introduce some notations required in the following proof. Let the row support set of $\B^k$ in the $k$ iteration be 
\begin{align*}
    A^k = \{1\leq i \leq p: \norm{\b_i^k}_2 \neq 0\},
\end{align*}
where $\b_i^k$ is the $i$th row of $\B^k$. We define the following index sets  
\begin{align}\label{def:set:1}
    A_{1}^k = A^k \cap A^*, \quad 
    A_{2}^k = A^* / A^k, \quad 
    I_{1}^k = A^k \cap I^*, \quad 
    I_{2}^k = I^* / A^k,
\end{align}
with $I^* = (A^*)^c$. Here $A_1^k$ is the true features selected in the $k$th iteration, and $A_2^k$ is the missed features in the $k$th iteration. Oppositely, $I_1^k$ is the false features selected in the $k$ step. $I_2^k$ is the unimportant features and not selected in the $k$th iteration. 

Similar to the above definition, there are corresponding sets from $k$ to $k+1$
\begin{align}\label{def:set:2}
    A_{11}^{k} = A_1^k / (A^{k+1} \cap A_1^k),~
    A_{22}^{k} = A_2^k / (A^{k+1} \cap A_2^k), ~ 
    I_{11}^{k} = A^{k+1} \cap I_1^k, ~ 
    I_{22}^k = A^{k+1} \cap I_{2}^k,
\end{align}
where $A_{11}^k$ is the missed trues features in $A_1^k$, and $A_{22}^k$ are missed true features in $A_2^k$. Oppositely, $I_{11}^k$ is the false features selected from $I_1^k$. $I_{22}^k$ is the false features selected from $I_2^k$. Moreover, we define the difference between the active set $A^k$ and the true support set $A^*$ as follows 
\begin{align*}
    D(A^k) = \norm{\C^*_{(A^*/A^k)\cdot}}_{\op} = \norm{\C^*_{A_2^k\cdot}}_{\op}, 
\end{align*}
which will be used in the following proofs. For convenience, we also define 
\begin{align}
    \bDelta^k 
    = \C_{A^k\cdot}^{k+1} - \C^*_{A^k\cdot},~ 
    \wt{\bGamma}_{A^k\cdot} = n^{-1}\X_{\cdot A^k}\trans (\Y - \X\C^{k+1}), ~
    \wt{\bGamma}_{I^k\cdot} = \0_{(p-s)\times q}, 
    \label{def:noise}
\end{align}
with $\C^{k+1} = \B^{k+1}(\V^{k+1})\trans$, which will be used later.  

\input{proof/main-result.tex}

\input{proof/proof-gic.tex}

\input{proof/proof-gic-2.tex}

\section{Auxiliary Lemmas}

\input{proof/lemma.tex}

%% file: proof/main-result.tex
\subsection{Proofs of Theorem \ref{theorem:1}}

The proof of Theorem \ref{theorem:1} relies on the complicated mathematical derivation, thus we divide complicated derivations into Lemmas \ref{lemma:1}--\ref{lemma:5}. The key of proof is to suppress the loss due to the missed true features. Combining \eqref{lem:2:ineq:1}--\eqref{lem:2:ineq:3}, we can get 
\begin{align}
    D(A^{k+1}) 
    &\leq \norm{\bDelta_{A_{11}^k\cdot}}_{\op} + \norm{\C^{k+1}_{A_{11}^k\cdot}}_{\op} 
    + \frac{\norm{\wt{\bGamma}^{k+1}_{A_{22}^k\cdot}}_{\op} + \theta_{s}\norm{\bDelta^k}_{\op} + \theta_{s}D(A^k) + h(s)}{c_{-}(s)} 
    \nonumber \\ 
    &\leq \frac{
        \norm{\bDelta_{A_{11}^k\cdot}}_{\op} 
        + \norm{\C^{k+1}_{A_{11}^k\cdot}}_{\op} 
        + \norm{\wt{\bGamma}^{k+1}_{A_{22}^k\cdot}}_{\op} 
        + \theta_{s}\norm{\bDelta^k}_{\op} 
        + \theta_{s}D(A^k) + h(s)} {c_{-}(s)} \label{ineq:1:th:1}
\end{align}
where we use the fact $0 < c_{-}(s) \leq 1$ because the each column of $\X$ is normalized to $\sqrt{n}$. Then applying Lemma \ref{lemma:4}, we have $\norm{\C^{k+1}_{A_{11}^k\cdot}}_{\op} = \norm{\B^{k+1}_{A_{11}^k\cdot}}_{\op}$. Thus utilizing Lemma \ref{lemma:5} yields 
\begin{align}
    \norm{\C^{k+1}_{A_{11}^k\cdot}}_{\op} + \norm{\wt{\bGamma}^{k+1}_{A_{22}^k\cdot}}_{\op} 
    &\leq 
    \norm{\B^{k+1}_{A_{11}^k\cdot}}_{\op} 
    + \norm{\bGamma^{k+1}_{A_{22}^k\cdot}}_{\op} 
    + c h(s) 
    \nonumber \\ 
    &\leq 
    \sqrt{r} (\norm{\C^{k+1}_{I_{11}^k}}_{\op} + \norm{\wt{\bGamma}^{k+1}_{I_{22}^k}}_{\op}) 
    + c h(s) \label{ineq:2:th:1}
\end{align}
where $c$ is a constant and the last inequality holds because of Lemma \ref{lemma:3}. 

Moreover, following the definition of $\wt{\bGamma}^{k+1}$ and similar simplifications in Lemma \ref{lemma:2}, we have 
\begin{align}
    \norm{\wt{\bGamma}^{k+1}_{I_{22}^k\cdot}}_{\op} 
    &= n^{-1}\norm{\X_{\cdot I_{22}^k}\trans\left(\Y - \X\C^{k+1}\right)}_{\op} 
    \leq n^{-1}\norm{\X_{\cdot I_{22}^k}\trans(\X_{\cdot A_2^k}\C^*_{A_2^k\cdot} + \E - \X_{\cdot A^k}\bDelta^k)}_{\op} 
    \nonumber \\
    &\leq \theta_{s} D(A^k) + \theta_{s}\norm{\bDelta^k}_{\op} + h(s), \label{ineq:3:th:1}
\end{align}
where $\C^{k+1} = \B^{k+1}(\V^{k+1})\trans$, $h(s) = \max_{\abs{A}\leq s}n^{-1}\norm{\X_{\cdot A}\trans\E}_{\op}$, and the last inequality holds due to Condition \ref{cond:2} and the triangle inequality. 

Then by the inequalities \eqref{ineq:2:th:1} and \eqref{ineq:3:th:1}, we have 
\begin{align}
    \norm{\C_{A_{11}^k\cdot}^{k+1}}_{\op} + \norm{\wt{\bGamma}_{A_{22}^k}^{k+1}}_{\op} 
    \leq \sqrt{r}\left(\norm{\bDelta^k}_{\op} 
    + \theta_{s}D(A^k) 
    + \theta_{s}\norm{\bDelta^k}_{\op} + h(s)\right) + c h(s), \label{ineq:4:th:1}
\end{align}
where we use the fact $\norm{\C_{I_{11}^k}^{k+1}}_{\op} \leq \norm{\bDelta^k}_{\op}$ with $\bDelta^k = \C^{k+1}_{A^k\cdot} - \C^*_{A^k\cdot}$ and $I_{11}^k\subset A^k$. 

Combining \eqref{ineq:4:th:1} and \eqref{ineq:1:th:1} yields 
\begin{align}
    D(A^{k+1}) \leq 
    \frac{(1+\sqrt{r})(1+\theta_{s})\norm{\bDelta^k}_{\op} + (1+\sqrt{r})\theta_{s}D(A^k) + (c+1+\sqrt{r}) h(s)}{c_{-}(s)}, 
    \label{ineq:5:th:1}
\end{align}
where we use the fact fact $\norm{\bDelta_{A_{11}^k}^{k+1}}_{\op} \leq \norm{\bDelta^k}_{\op}$. Then applying the inequality \eqref{lemma:1:ineq:1} in Lemma \ref{lemma:1}, we can obtain 
\begin{align}
    \norm{\bDelta^k}_{\op} 
    = \norm{\C^{k+1}_{A^k\cdot} - \C^*_{A^k\cdot}}_{\op} 
    \leq \frac{\theta_{s}}{c_{-}(s)} D(A^k) 
    + \frac{2h(s)}{c_{-}(s)}. \label{ineq:6:th:1}
\end{align}
Note that there exists a large enough constant $c^{\prime}$, which is subject to $c+1+\sqrt{r} \leq c^{\prime}(1+\sqrt{r})$. Then after combining \eqref{ineq:5:th:1} and \eqref{ineq:6:th:1}, some algebraic simplifications yields 
\begin{align*}
    D(A^{k+1}) 
    &\leq 
    \left[\frac{\theta_{s}(1+\sqrt{r})(1+\theta_{s})}{c_{-}^2(s)} + \frac{(1+\sqrt{r})\theta_{s}}{c_{-}(s)}\right] D(A^k) 
    \\
    &\quad~ + c^{\prime} \left[\frac{(1+\sqrt{r})(1+\theta_{s})}{c_{-}^2(s)} + \frac{1+\sqrt{r}}{c_{-}(s)}\right] h(s), 
\end{align*}
where $c^{\prime}$ is a positive constant. Then define the $\gamma$ as follows 
\begin{align*}
    \gamma 
    = \frac{\theta_{s}(1+\sqrt{r})(1+\theta_{s})}{c_{-}^2(s)} + \frac{(1+\sqrt{r})\theta_{s}}{c_{-}(s)},
\end{align*}
which ensures that $D(A^{k+1}) \leq \gamma D(A^k) + c^{\prime}\frac{\gamma}{\theta_{s}} h(s)$. Therefore, applying the inequality recursively and supposing $\gamma < 1$, we have 
\begin{align*}
    D(A^{k+1}) 
    &\leq \gamma^{k+1} D(A^0) + (1 + \gamma + \cdots + \gamma^{k+1}) \frac{c^{\prime} h(s)}{\theta_{s}} 
    \\
    &\leq \gamma^{k+1} \norm{\C^*}_{\op} + c^{\prime}\frac{\gamma}{\theta_{s}(1-\gamma)} h(s),
\end{align*}
where $c^{\prime}$ is a positive constant. Then by the inequality \eqref{lemma:1:ineq:2}, there is 
\begin{align}
    \norm{\C^{k+1} - \C^*}_{\op} 
    &\leq \left(1 + \frac{\theta_{s}}{c_{-}(s)}\right) 
    \left(\gamma^k \norm{\C^*}_{\op} + c\frac{\gamma}{\theta_{s}(1-\gamma)} h(s)\right) 
    + \frac{2h(s)}{c_{-}(s)} 
    \nonumber \\
    &\leq \gamma^k \left(1 + \frac{\theta_{s}}{c_{-}(s)}\right) \norm{\C^*}_{\op} + C h(s), 
    \label{ineq:7:th:1}
\end{align}
where $c$ and $C$ are some positive constants. Recall the definition of $h(s)$ as follows 
\begin{align*}
    h(s) = n^{-1}\max_{\abs{A}\leq s}~\norm{\X_{\cdot A}\trans\E}_{\op},
\end{align*}
then applying Lemma \ref{lemma:10}, we have 
\begin{align}
    h(s) \leq O\left(n^{-1/2}\{s + q + s\log(ep/s)\}^{1/2}\right) \label{ineq:8:th:1}
\end{align}
holds with the probability at least $1 - p^{-c}$. Finally, note that $\norm{\C^{k+1}-\C^*}_{F}\leq \sqrt{2r} \norm{\C^{k+1}-\C^*}_{\op}$, and then combining \eqref{ineq:7:th:1} and \eqref{ineq:8:th:1} will conclude the proof.

%% file: proof/proof-gic.tex
\subsection{Proofs of Theorem \ref{theorem:2}}

Denote by $\wh{\C}$ the estimation of MrBeSS with the rank $r$ and row support set $A$. Note that $\wh{\C}$ actually is the reduced rank estimator constrained in the support set $A$ and $\rank{\wh{\C}}=r$. For the sake of clarity, we first define the event as follows and then prove the result conditional on the event 
\begin{align*}
    \mathcal{G}(\bar{s}) = \left\{n^{-1}\left(\norm{\E\trans\X_{\cdot J}}_{\op}
    \vee \norm{\E\trans\X_{\cdot A}}_{\op} \vee \norm{\E\trans\X_{\cdot A^*}}_{\op}\right) \leq C\tau_n(\bar{s})\right\},
\end{align*}
with $J = A^* / A$, $\bar{s}=\max\{\abs{A},\abs{A^*}\}$ and $\tau_n(s)=n^{-1/2}\{s+q+s\log(ep/s)\}^{1/2}$. Then there are two kinds of cases as follows. 

\noindent\underline{1. Underfitted case: $A^*\not\subset A$ or $r < r^*$.} Define the signal strength parameter as follows 
\begin{align*}
    \delta_n = \big\{\underset{A^*\not\subset A}{\inf}~n^{-1}\norm{(\I - \bP_A)\X\C^*}_F^2\big\} \wedge \sigma_{r^*}^*,
\end{align*}
where $\sigma_{r^*}^*$ is the $r^*$th largest eigenvalue of $n^{-1}\C\strans\X\trans\X\C^*$. Therefore from Lemma \ref{lemma:8}, we have 
\begin{align*}
    \left(L(\wh{\C}) - L(\C^*)\right) 
    \geq \delta_n - c \bar{s}\tau_n(\bar{s})\norm{\C^*}_{\op},
\end{align*}
with the underfitted case and some positive constant $c$. Furthermore, combining Lemma \ref{lemma:9} and the above inequality yields 
\begin{align}
    \left(L(\wh{\C}) - L(\wh{\C}^{\ora})\right) 
    \geq \delta_n - C (\bar{s} \vee r^*) \tau_n(\bar{s})\norm{\C^*}_{\op}, 
    \label{ineq:1:th:2}
\end{align}
under the event $\mathcal{G}(\bar{s})$, where $\wh{\C}^{\ora}$ is the estimator of the oracle reduced rank regression with row support set $A^*$ and rank $r^*$. 

\bigskip 

\noindent\underline{2. Overfitted case: $A^*\subsetneq A$ and $r > r^*$.} Similar to the above argument, utilizing Lemma \ref{lemma:9} and \ref{lemma:8}, we can obtain  that 
\begin{align}
    \abs{L(\wh{\C}) - L(\wh{\C})^{\ora}} 
    \leq C (\bar{s} \vee r^*) \tau_n(\bar{s})\norm{\C^*}_{\op}, 
    \label{ineq:3:th:2}
\end{align}
under the event $\mathcal{G}(\bar{s})$. 

Then recall the definition of the generalized information criterion (GIC) as follows 
\begin{align*}
    \GIC(\C) 
    = L(\C) + (s+r) a_n, 
\end{align*}
where $s$ denotes the number of the nonzero rows in $\C$ and $r=\rank{\C}$. Thus for any estimator $\wh{\C}$ with row support set $\abs{A}=s$ and $\rank{\wh{\C}}=r$, there is 
\begin{align}
    \GIC(\wh{\C}) - \GIC(\wh{\C}^{\ora}) 
    = L(\wh{\C}) - L(\wh{\C}^{\ora}) 
    + (s+r-s^*-r^*) a_n. 
    \label{ineq:2:th:2}
\end{align}
With the underfitted $\wh{\C}$, combining \eqref{ineq:1:th:2} and \eqref{ineq:2:th:2} yields 
\begin{align}
    &\quad~ \GIC(\wh{\C}) - \GIC(\wh{\C}^{\ora}) 
    \nonumber \\
    &\geq \delta_n - C(\bar{s}\vee r^*)\tau_n(\bar{s}) \norm{\C^*}_{\op}
    + (s+r-s^*-r^*) a_n 
    \nonumber \\ 
    &\geq \delta_n - C(\bar{s}\vee r^*)\tau_n(\bar{s}) \norm{\C^*}_{\op}
    - \abs{s+r-s^*-r^*} a_n. \label{ineq:5:th:2}
\end{align}

Similarly, we compare the difference between $\wh{\C}$ and $\wh{\C}^{\ora}$ under the overfitted case. And then by the inequality \eqref{ineq:3:th:2}, we have 
\begin{align}
    L(\wh{\C}) - L(\wh{\C}^{\ora}) 
    \geq - C(\bar{s}\vee r^*)\tau_n(\bar{s}) \norm{\C^*}_{\op}. 
    \label{ineq:4:th:2}
\end{align}
Furthermore combining \eqref{ineq:2:th:2} and \eqref{ineq:4:th:2}, we can get 
\begin{align*}
    \GIC(\wh{\C}) - \GIC(\wh{\C}^{\ora}) 
    &\geq (s+r-s^*-r^*) a_n
    - C(\bar{s}\vee r^*)\tau_n(\bar{s}) \norm{\C^*}_{\op}. 
\end{align*}
Note that under the overfitted case, we have $s>s^*$ and $r>r^*$. Therefore, it yields 
\begin{align}
    \GIC(\wh{\C}) - \GIC(\wh{\C}^{\ora}) 
    \geq 2 a_n 
    - C(\bar{s}\vee r^*)\tau_n(\bar{s}) \norm{\C^*}_{\op},
    \label{ineq:6:th:2}
\end{align}
under the overfitted case. 

Assume that we tune the $s$ and $r$ in the parameter space as follows 
\begin{align*}
    \mathcal{C} = \left\{\C\in\mbR^{p\times q}:\text{the nonzero rows of $\C\leq s$, $\rank{\C}\leq r$ and $r\leq s \leq K$}\right\},
\end{align*}
with some $K \geq \bar{s} \geq r^*$. Thus with the underfitted $\wh{\C}$, we have 
\begin{align*}
    \GIC(\wh{\C}) - \GIC(\wh{\C}^{\ora}) 
    \geq \delta_n - C K \tau_n(K)\norm{\C^*}_{\op} - 4K a_n,
\end{align*}
by the inequality \eqref{ineq:5:th:2}. Similar to the above and with the overfitted $\wh{\C}$, there is 
\begin{align*}
    \GIC(\wh{\C}) - \GIC(\wh{\C}^{\ora}) 
    \geq 2 a_n - C K \tau_n(K)\norm{\C^*}_{\op},  
\end{align*}
from the inequality \eqref{ineq:6:th:2}.

If the following assumptions hold simultaneously 
\begin{align*}
    &K\tau_n(K)\norm{\C^*}_{op} = o(\delta_n), 
    \\
    &K a_n = o(\delta_n), 
    \\ 
    &K\tau_n(K)\norm{\C^*}_{op} = o\left(a_n\right), 
\end{align*}
we can obtain that 
\begin{align*}
    \underset{A\neq A^*~\text{or}~r\neq r^*}{\inf}\GIC(\wh{\C}) > \GIC(\wh{\C}^{\ora}),~\text{under the event}~\mathcal{G}(\bar{s}),
\end{align*}
where $A$ is the row support set of $\wh{\C}$ and $\rank{\wh{\C}} = r$. Specifically, we set $a_n = \{n^{-1}\log p\}^{1/2}q\log\log n$ then the above three assumptions can be concluded from 
\begin{align*}
    &K\{n^{-1}(q + K\log p)\}^{1/2}\norm{\C^*}_{\op} = o(\delta_n), 
    \\ 
    &K q(\log\log n)\{n^{-1}\log p\}^{1/2} = o(\delta_n),
    \\
    & K(\log p)^{-1/2}\norm{\C^*}_{\op} = o(q^{1/2}\log\log n),
\end{align*}
where we use the fact $K \ll q$. Finally, applying Lemma \ref{lemma:10} yields that $\mathcal{G}(K)$ holds with the probability at least $1-p^{-c}$, which finishes the proof. 

%% file: proof/proof-gic-2.tex
\subsection{Proofs of Theorem \ref{theorem:3}}

In this section, we denote the output of MrBeSS by $\C$ for the sake of clarity. With the given maximum rank $r_{\max} > r^*$, we first show that
\begin{align*}
    \wh{\C} = \underset{\C\in \mathcal{C}(K)}{\arg\min} ~ \text{GIC}(\C) \quad~  \text{s.t.}~ \rank{\C} =r_{\max},
\end{align*}
leads to $\wh{A}=A^*$ with high probability, where $\wh{A}$ is the row support set of $\wh{\C}$. Then under the given support set $A^*$, we can recover the true rank by minimizing GIC with high probability. Similar to the proof of Theorem \ref{theorem:2}, we mainly prove the result conditional on the event $\mathcal{G}(K)$, where $K$ is the maximum level of the search space. Moreover, denote the number of nonzero rows in $\C$ by $s$  and $\tau_n(K) = \{(K+q)n^{-1}\log p\}^{1/2}$ for convenience. 

With the given rank $r_{\max}$, the GIC defined in Theorem \ref{theorem:2} is 
\begin{align*}
    \text{GIC}(\C) = L(\C) + q(s+r_{\max})(\log\log n)\{n^{-1}\log p\}^{1/2},
\end{align*}
with $L(\C) = (2n)^{-1}\norm{\Y - \X\C}_F^2$. Under the assumption $r_{\max} > r^*$ and applying Case 1 and 2 in Lemma \ref{lemma:8}, we can obtain a similar result in the proof of Theorem \ref{theorem:2}, which is as follows. On the one hand, we have
\begin{align*}
    \text{GIC}(\C) - \text{GIC}(\C^*)
     & = L(\C) - L(\C^*) + q(s - s^*)(\log\log n)\{n^{-1}\log p\}^{1/2}
    \nonumber                                                                                           \\
     & \geq (2n)^{-1}\norm{(\I-\bP_{A})\X\C^*}_F^2 - C \bar{s}\tau_n(\bar{s})\norm{\C^*}_{\op}
    \nonumber                                                                                           \\
     & \quad~  - \abs{s - s^*}q(\log\log n)\{n^{-1}\log p\}^{1/2},
\end{align*}
with $A\not\subset A^*$, $\rank{\C}=r_{\max}$ and $\bar{s} = \max\{\abs{A},\abs{A^*/A}\}$. On the other hand, there is
\begin{align*}
    \text{GIC}(\C) - \text{GIC}(\C^*)
    \geq \abs{s - s^*}q(\log\log n)\{n^{-1}\log p\}^{1/2}
    - C \bar{s} \tau_n(\bar{s})\norm{\C^*}_{\op},
\end{align*}
with $A^*\subsetneq A$ and $\rank{\C}=r_{\max}$. By Lemma \ref{lemma:9} and the above two inequalities, under the event $\mathcal{G}(K)$, we can get

\noindent\underline{1) $A\not\subset A^*$ and $\rank{\C}=r_{\max}$.}
\begin{align*}
    \text{GIC}(\C) - \text{GIC}(\wt{\C}^{\ora})
     & \geq \delta_n - \abs{s-s^*} q (\log\log n)\{n^{-1}\log p\}^{1/2}
    \nonumber                                                                       \\
     & \quad~ - C (\bar{s}\vee r_{\max})\{n^{-1}(\bar{s} + q)\log p\}^{1/2}\norm{\C^*}_{\op}.
\end{align*}

\noindent\underline{2) $A\subsetneq A^*$ and $\rank{\C}=r_{\max}$.}
\begin{align*}
    \text{GIC}(\C) - \text{GIC}(\wt{\C}^{\ora})
     & \geq (s-s^*)q(\log\log n)\{n^{-1}\log p\}^{1/2}
    \nonumber                                                       \\
     & \quad~ - C (\bar{s}\vee r_{\max})\{n^{-1}(\bar{s} + q)\log p\}^{1/2}\norm{\C^*}_{\op},
\end{align*}
where the row support set of $\wt{\C}^{\ora}$ is $A^*$ but $\rank{\wt{\C}^{\ora}}=r_{\max}$. Note that we search the solution with $\max\,\{s,r_{\max},s^*,r^*\}\leq K$ in the parameter space $\mathcal{C}$. By the similar argument in the proof of Theorem \ref{theorem:2}, we have
\begin{align*}
    &\text{GIC}(\C) - \text{GIC}(\wt{\C}^{\ora})
    \\
    &\,\,\, \geq \delta_n - C K\tau_n(K)\norm{\C^*}_{\op}
     - 2Kq(\log\log n) \{n^{-1}\log p\}^{1/2}, 
     \,\, \text{if}~ A\not\subset A^*, 
     \\
    &\text{GIC}(\C) - \text{GIC}(\wt{\C}^{\ora}) 
    \\
    &\,\,\, \geq q(\log\log n) \{n^{-1}\log p\}^{1/2} - C K \tau_n(K)\norm{\C^*}_{\op}, 
    \,\, \text{if}~ A\subsetneq A^*,
\end{align*}
with $\rank{\C}=\rank{\wt{\C}^{\ora}} = r_{\max}$. Thus under the same assumptions in Theorem \ref{theorem:2} and conditional on the event $\mathcal{G}(K)$, there is 
\begin{align*}
    \text{GIC}(\C) > \text{GIC}(\wt{\C}^{\ora}),~\text{with}~A\neq A^*~\text{and large enough}~n, 
\end{align*}
constrained in $\rank{\C}=\rank{\wt{\C}^{\ora}}=r_{\max}$. 


Next we show that with the given $A^*$, minimizing the GIC can recover the true rank of $\C^*$ conditional on the event $\mathcal{G}(K)$ and with the large enough $n$. Directly applying Cases 1 -- 4 in Lemma \ref{lemma:8} and by the similar argument in the proof of Theorem \ref{theorem:2}, we can obtain that 
\begin{align*}
    \text{GIC}(\C) - \text{GIC}(\wh{\C}^{\ora}) \geq \delta_n - C K \tau_n(K)\norm{\C^*}_{\op} - 2K q\{n^{-1}\log p\}^{1/2} \log\log n,
\end{align*}
with the underfitted estimator $\C$. On the other hand, with the overfitted $\C$, there is 
\begin{align*}
    \text{GIC}(\C) - \text{GIC}(\wh{\C}^{\ora}) 
    \geq \{n^{-1}\log p\}^{1/2} q \log\log n  - C K \tau_n(K)\norm{\C^*}_{\op}, 
\end{align*}
where $\wh{\C}^{\ora}$ is the oracle reduced rank estimator with $\rank{\wh{\C}^{\ora}}=r^*$ and the support set $A^*$. Thus under the same assumptions of Theorem \ref{theorem:2} and conditional on $\mathcal{G}(K)$, we can get 
\begin{align*}
    \text{GIC}(\C) > \text{GIC}(\wh{\C}^{\ora}) \,\,\text{with the large enough}\,\, n.
\end{align*}
Finally applying Lemma \ref{lemma:10}, we have $\mathcal{G}(K)$ holds with the probability at least $1 - p^{-c}$ that concludes the proof.  

%% file: proof/lemma.tex
\subsection{Lemma \ref{lemma:1} and its proofs}

\begin{lemma}\label{lemma:1}
    Assume that Conditions \ref{cond:1} and \ref{cond:2} hold. Then we have
    \begin{align}
         & \norm{\C^{k+1}_{A^k\cdot} - \C^*_{A^k\cdot}}_{\op}
        \leq \frac{\theta_{s}}{c_{-}(s)} D(A^k) 
        + \frac{2h(s)}{c_{-}(s)} ,
        \label{lemma:1:ineq:1}
        \\
         & \norm{\C^{k+1} - \C^*}_{\op}. \label{lemma:1:ineq:2}
        \leq \left(1 + \frac{\theta_{s}}{c_{-}(s)}\right) D(A^k) 
        + \frac{2h(s)}{c_{-}(s)},
    \end{align}
    where $\C^{k+1} = \B^{k+1}(\V^{k+1})\trans$ and $h(s) = \max_{\abs{A}\leq s} n^{-1}\norm{\X_{\cdot A}\trans\E}_{\op}$.
\end{lemma}

Following the algorithm of MrBeSS, we have
\begin{align*}
    \B_{A^k\cdot}^{k+1} = (\X_{\cdot A^k}\trans\X_{\cdot A^k})^{-1}\X_{\cdot A^k}\Y\V^{k+1},
\end{align*}
after updating $\V^{k+1}$. For the sake of clarity, we denote $\B^{k+1}(\V^{k+1})\trans$ by $\C^{k+1}$. Therefore, we have
\begin{align*}
    \C^{k+1}_{A^k\cdot}
    = (\X_{\cdot A^k}\trans\X_{\cdot A^k})^{-1}\X_{\cdot A^k}\trans\Y\V^{k+1}(\V^{k+1})\trans.
\end{align*}
Moreover, there is a fact $\C_{A^k\cdot}^* = (\X_{\cdot A^k}\trans\X_{\cdot A^k})^{-1} \X_{\cdot A^k}\trans\X_{\cdot A^k} \C_{A^k\cdot}^*$. It yields
\begin{align*}
    \C_{A^k\cdot}^*
    = (\X_{\cdot A^k}\trans\X_{\cdot A^k})^{-1} \X_{\cdot A^k}\trans\Y
    - \bXi 
    - (\X_{\cdot A^k}\trans\X_{\cdot A^k})^{-1} \X_{\cdot A^k}\trans\X_{\cdot A_2^k}\C^*_{A_2^k\cdot},
\end{align*}
where $\bXi=(\X_{\cdot A^k}\trans\X_{\cdot A^k})^{-1} \X_{\cdot A^k}\trans\E$, and we use the fact $\Y=\X_{\cdot A^k}\C^*_{A^k\cdot} + \X_{\cdot A_2^k}\C^*_{A_2^k\cdot} + \E$. Because the rows of $\C^*_{A^k\cdot}$ restricted in $A^k/A_1^k$ are zeros, and $A_1^k\cup A_2^k = A^*$. Recall the definition of $\V^{k+1}$, and they are also the top-$r$ right singular vectors of
\begin{align*}
    (\X_{\cdot A^k}\trans\X_{\cdot A^k})^{-1/2} \X_{\cdot A^k}\trans\Y = \sum_{\ell=1}^{q} d_{\ell}^{k+1}\u_{\ell}^{k+1}(\v_{\ell}^{k+1})\trans,
\end{align*}
where $d_{\ell}^{k+1}$ and $\u_{\ell}^{k+1}$ are the $\ell$th largest singular value and its corresponding left singular vector, respectively. Then there is
\begin{align*}
    \C_{A^k\cdot}^{k+1} - \C_{A^k\cdot}^*
     & = -(\X_{\cdot A^k}\trans\X_{\cdot A^k})^{-1/2}\sum_{\ell=r+1}^q d_{\ell}^{k+1}\u_{\ell}^{k+1}\v_{\ell}^{k+1}
    \nonumber                                                                                                       \\
     & \quad~ + \bXi
    + (\X_{\cdot A^k}\trans\X_{\cdot A^k})^{-1} \X_{\cdot A^k}\trans\X_{\cdot A_2^k}\C^*_{A_2^k\cdot}.
\end{align*}
By the above equation, we can obtain
\begin{align}
    \norm{\C_{A^k\cdot}^{k+1} - \C_{A^k\cdot}^* }_{\op}
     & \leq d_{r+1}^{k+1}\norm{(\X_{\cdot A^k}\trans\X_{\cdot A^k})^{-1/2}}_{\op}
    + \norm{\bXi}_{\op}
    + \norm{(\X_{\cdot A^k}\trans\X_{\cdot A^k})^{-1} \X_{\cdot A^k}\trans\X_{\cdot A_2^k}\C_{A_2^k\cdot}}_{\op}
    \nonumber                                                                     \\
     & \leq d_{r+1}^{k+1}\norm{(\X_{\cdot A^k}\trans\X_{\cdot A^k})^{-1/2}}_{\op}
    + \norm{\bXi}_{\op}
    + c_{-}^{-1}(s)\theta_{s} \norm{\C^*_{A_2^k\cdot}}_{\op}, \label{ineq:1:lem:1}
\end{align}
where the last inequality is from Conditions \ref{cond:1} and \ref{cond:2}.

In addition, note that $d_{r+1}^{k+1}$ is the $(r+1)$th largest singular value of
\begin{align*}
    (\X_{\cdot A^k}\trans\X_{\cdot A^k})^{-1/2} \X_{\cdot A^k}\trans\Y
    = (\X_{\cdot A^k}\trans\X_{\cdot A^k})^{-1/2} \X_{\cdot A^k}\trans\X\C^*
    + (\X_{\cdot A^k}\trans\X_{\cdot A^k})^{-1/2} \X_{\cdot A^k}\trans\E,
\end{align*}
and $\rank{\C^*} = r^* \leq r$. Then by the matrix perturbation theory, we have
\begin{align}
    d_{r+1}^{k+1}\cdot \norm{(\X_{\cdot A^k}\trans\X_{\cdot A^k})^{-1/2}}_{\op}
     & \leq \norm{(\X_{\cdot A^k}\trans\X_{\cdot A^k})^{-1/2} \X_{\cdot A^k}\trans\E}_{\op}
    \cdot \norm{(\X_{\cdot A^k}\trans\X_{\cdot A^k})^{-1/2}}_{\op}
    \nonumber                                                                                                \\
     & \leq \norm{(\X_{\cdot A^k}\trans\X_{\cdot A^k})^{-1}}_{\op} \cdot \norm{\X_{\cdot A^k}\trans\E}_{\op}
    \nonumber                                                                                                \\
     & \leq c_{-}^{-1}(s)\cdot n^{-1}\norm{\X_{\cdot A^k}\trans\E}_{\op}, \label{ineq:2:lem:1}
\end{align}
where the last inequality is due to Condition \ref{cond:1}. Similarly, recall the definition of $\bXi$, and there is
\begin{align}
    \norm{\bXi}_{\op}
    \leq \norm{(\X_{\cdot A^k}\trans\X_{\cdot A^k})^{-1}}_{\op} \cdot \norm{\X_{\cdot A^k}\trans\E}_{\op}
    \leq c_{-}^{-1}(s)\cdot n^{-1}\norm{\X_{\cdot A^k}\trans\E}_{\op}. \label{ineq:3:lem:1}
\end{align}
Thus combining \eqref{ineq:1:lem:1}--\eqref{ineq:3:lem:1}, we will obtain
\begin{align*}
    \norm{\C^{k+1}_{A^k\cdot} - \C^*_{A^k\cdot}}_{\op}
    \leq c_{-}^{-1}(s)\theta_{s}\norm{\C^*_{A_2^k\cdot}}_{\op}
    + 2 c_{-}^{-1}(s)\cdot n^{-1}\norm{\X_{\cdot A^k}\trans\E}_{\op}.
\end{align*}
As for the upper bound of $\norm{\C^{k+1} - \C^*}_{\op}$, we have
\begin{align*}
    \norm{\C^{k+1} - \C^*}_{\op}
     & \leq \norm{\C_{A^k\cdot}^{k+1} - \C_{A^k\cdot}^*}_{\op}
    + \norm{\C_{A_2^k\cdot}^*}_{\op}
    \nonumber                                                                            \\
     & \leq \left(1 + \frac{\theta_{s}}{c_{-}(s)}\right)\norm{\C^*_{A_2^k\cdot}}_{\op}
    + 2 c_{-}^{-1}(s)\cdot n^{-1}\norm{\X_{\cdot A^k}\trans\E}_{\op}.
\end{align*}
Finally, define $h(s) = \max_{\abs{A}\leq s} n^{-1}\norm{\X_{\cdot A}\trans\E}_{\op}$, then it finishes the proof.

\subsection{Lemma \ref{lemma:2} and its proofs}

\begin{lemma}\label{lemma:2}
    With the definitions in \eqref{def:set:1} and \eqref{def:set:2}, there are
    \begin{align}
         & D(A^{k+1})
        \leq \norm{\C^*_{A_{11}^k\cdot}}_{\op}
        + \norm{\C^*_{A_{22}^k\cdot}}_{\op}, \label{lem:2:ineq:1}
        \\
         & \norm{\C^*_{A_{11}^k\cdot}}_{\op}
        \leq \norm{\bDelta_{A_{11}^k\cdot}}_{\op} + \norm{\C^{k+1}_{A_{11}^k\cdot}}_{\op}. \label{lem:2:ineq:2}
    \end{align}
    Furthermore, assume that Conditions \ref{cond:1} and \ref{cond:2} hold. We have
    \begin{align}\label{lem:2:ineq:3}
        \norm{\C^*_{A_{22}^k}}_{\op}
        \leq \frac{\norm{\wt{\bGamma}^{k+1}_{A_{22}^k\cdot}}_{\op} + \theta_{s}\norm{\bDelta^k}_{\op} + \theta_{s}D(A^k) + h(s)}{c_{-}(s)}
    \end{align}
\end{lemma}

By the definitions of $A_{11}^k$ and $A_{22}^k$, we have $A^*/A^{k+1} = A_{11}^k \cup A_{22}^k$ and $A_{11}^k \cap A_{22}^k = \varnothing$. On the one hand, the triangle inequality yields
\begin{align*}
    D(A^{k+1})
    = \norm{\C^*_{A^*/A^{k+1}}}_{\op}
    \leq \norm{\C^*_{A_{11}^k\cdot}}_{\op}
    + \norm{\C^*_{A_{22}^k\cdot}}_{\op}.
\end{align*}
On the other hand, note that $\bDelta^k = \C^{k+1}_{A^k\cdot} - \C^*_{A^k\cdot}$. Applying the triangle inequality again, we have
\begin{align*}
    \norm{\C^{k+1}_{A_{11}^k\cdot}}_{\op}
    = \norm{\C^*_{A_{11}^k\cdot} + \bDelta_{A_{11}^k\cdot}}_{\op}
    \geq \norm{\C^*_{A_{11}^k\cdot}}_{\op} - \norm{\bDelta_{A_{11}^k\cdot}}_{\op}.
\end{align*}
Then we get the inequalities in \eqref{lem:2:ineq:1} and \eqref{lem:2:ineq:2}.

Next we prove the inequality \eqref{lem:2:ineq:3}. Following the definition of $\wt{\bGamma}^{k+1}$, we have
\begin{align}
    \norm{\wt{\bGamma}^{k+1}_{A_{22}^k\cdot}}_{\op}
     & = n^{-1}\norm{\X_{\cdot A_{22}^k}\trans\left(\Y - \X_{\cdot A^k}\B^{k+1}_{A^k\cdot}(\V^{k+1})\trans\right)}_{\op}
    \nonumber                                                                                                                  \\
     & = n^{-1}\norm{\X_{\cdot A_{22}^k}\trans\left(\Y - \X_{\cdot A^k}\C^{k+1}_{A^k\cdot}\right)}_{\op}. \label{ineq:1:lem:2}
\end{align}
Moreover, by some simplifications, there is
\begin{align}
    \Y
     & = \X\C^{k+1} + \X(\C^* - \C^{k+1}) + \E
    = \X_{\cdot A^k}\C^{k+1}_{A^k\cdot} - \X_{\cdot A^k}\bDelta^k + \X_{\cdot A_2^k} \C^*_{A_2^k} + \E
    \nonumber                                                        \\
     & = \X_{\cdot A^k}\C^{k+1}_{A^k\cdot} - \X_{\cdot A^k}\bDelta^k
    + \X_{\cdot A_{22}^k} \C^*_{A_{22}^k}
    + \X_{\cdot (A_2^k/A_{22}^k)} \C^*_{(A_2^k/A_{22}^k)\cdot}
    + \E. \label{ineq:2:lem:2}
\end{align}
Thus combining \eqref{ineq:1:lem:2} and \eqref{ineq:2:lem:2}, we will get
\begin{align*}
    \norm{\wt{\bGamma}^{k+1}_{A_{22}^k\cdot}}_{\op}
     & = n^{-1}\norm{
    \X_{\cdot A_{22}^k}\trans(\X_{\cdot A^k}\bDelta^k
    - \X_{\cdot A_{22}^k} \C^*_{A_{22}^k}
    - \X_{\cdot (A_2^k/A_{22}^k)} \C^*_{(A_2^k/A_{22}^k)\cdot}
    - \E)
    }_{\op}
    \\
     & \geq n^{-1}\norm{\X_{\cdot A_{22}^k}\trans\X_{\cdot A_{22}^k} \C^*_{A_{22}^k}}_{\op}
    - n^{-1}\norm{\X_{\cdot A_{22}^k}\trans\X_{\cdot A^k}\bDelta^k}_{\op}
    \\
     & \quad - n^{-1}\norm{\X_{\cdot A_{22}^k}\trans\X_{\cdot (A_2^k/A_{22}^k)} \C^*_{(A_2^k/A_{22}^k)\cdot}}_{\op}
    - n^{-1}\norm{\X_{\cdot A_{22}^k}\trans\E}_{\op}.
\end{align*}
By Conditions \ref{cond:1} and \ref{cond:2}, we have
\begin{align*}
    \norm{\wt{\bGamma}^{k+1}_{A_{22}^k\cdot}}_{\op}
    \geq c_{-}(s) \norm{\C^*_{A_{22}^k}}_{\op} - \theta_{s}\norm{\bDelta^k}_{\op} - \theta_{s}D(A^k) - h(s),
\end{align*}
where we use the fact $D(A^k) = \norm{\C^*_{A_2^k\cdot}}_{\op} \geq \norm{\C^*_{(A_2^k/A_{22}^k)\cdot}}_{\op}$ and $h(s) = \max_{\abs{A}\leq s}n^{-1}\norm{\X_{\cdot A}\trans\E}_{\op}$. Thus we finish the proof.

\subsection{Lemma \ref{lemma:3} and its proofs}

\begin{lemma}\label{lemma:3}
    Following the algorithm of MrBeSS, we have
    \begin{align*}
        \norm{\B_{A^k_{11}\cdot}^{k+1}}_{\op} + \norm{\bGamma_{A^k_{22}\cdot}^{k+1}}_{\op}
        \leq \sqrt{r} (\norm{\C_{I^k_{11}\cdot}^{k+1}}_{\op} + \norm{\wt{\bGamma}_{I^k_{22}\cdot}^{k+1}}_{\op}),
    \end{align*}
    with $\wt{\bGamma}_{I^k\cdot}^{k+1} = \X_{\cdot I^k}\trans(\Y - \X\B^{k+1}\V^{k+1}) / n$.
\end{lemma}

Recall that $\b_i^{k+1}$ and $\bGamma_i^{k+1}$ are the $i$th rows of $\B^{k+1}$ and $\bGamma^{k+1}$. And there are
\begin{align*}
    \bGamma_{I^k\cdot}^{k+1} = \X_{\cdot I^k}\trans(\Y\V^{k+1} - \X\B^{k+1}) / n, ~
    \bGamma_{A^k}^{k+1} = \0.
\end{align*}
Because $A_{11}^k$ and $A_{22}^k$ are the missed features in the iterative algorithm. Thus by the algorithm and the definitions of $A_{11}^k$ $A_{22}^k$, $I_{11}^k$ and $I_{22}^k$, the elements of $\{\norm{\b_i^{k+1}+\bgamma_i^{k+1}}_2: i\in A_{11}^k\cup A_{22}^k\}$ are no greater than the elements in $\{\norm{\b_i^{k+1}+\bgamma_i^{k+1}}_2: i\in I_{11}^k\cup I_{22}^k\}$. In addition, $\b_i^{k+1}$ and $\bgamma_i^{k+1}$ are complementary from \eqref{eq:kkt:approx}. Then there is
\begin{align*}
     & \max\{\norm{\b_i^{k+1}}_2: i\in A_{11}^k\} \vee \max\{\norm{\bgamma_i^{k+1}}_2: i\in A_{22}^k\}
    \nonumber                                                                                          \\
     & \quad~ \leq
    \min\{\norm{\b_i^{k+1}}_2: i\in I_{11}^k\} \wedge \min\{\norm{\bgamma_i^{k+1}}_2: i\in I_{22}^k\}.
\end{align*}

On the one hand, we can get
\begin{align}\label{ineq:1:lem:3}
    \norm{\B_{A^k_{11}\cdot}^{k+1}}_{\op} + \norm{\bGamma_{A^k_{22}\cdot}^{k+1}}_{\op}
    \leq \norm{\B_{A^k_{11}\cdot}^{k+1}}_{F} + \norm{\bGamma_{A^k_{22}\cdot}^{k+1}}_{F}
    \leq \norm{\B_{I^k_{11}\cdot}^{k+1}}_{F} + \norm{\bGamma_{I^k_{22}\cdot}^{k+1}}_{F}.
\end{align}
On the other hand, note that
\begin{align*}
    \bGamma_{I_k\cdot}^{k+1}
    = \X_{\cdot I^k}\trans(\Y\V^{k+1} - \X\B^{k+1}) / n
    = \X_{\cdot I^k}\trans(\Y - \X_{\cdot A^k}(\X_{\cdot A^k}\trans\X_{\cdot A^k})^{-1}\X_{\cdot A^k}\trans\Y)\V^{k+1} / n,
\end{align*}
which yields that $\rank{\bGamma^{k+1}} \leq r$. Then we have
\begin{align}\label{ineq:2:lem:3}
    \norm{\B_{I^k_{11}\cdot}^{k+1}}_{F} + \norm{\bGamma_{I^k_{22}\cdot}^{k+1}}_{F}
    \leq \sqrt{r} (\norm{\B_{I^k_{11}\cdot}^{k+1}}_{\op} + \norm{\bGamma_{I^k_{22}\cdot}^{k+1}}_{\op})
\end{align}

Next we will show the upper bound of $\norm{\bGamma_{I^k_{22}\cdot}^{k+1}}_{\op}$. For the sake of clarity, we denote that $\bP_{A^k} = \X_{\cdot A^k}(\X_{\cdot A^k}\trans\X_{\cdot A^k})^{-1}\X_{\cdot A^k}\trans$ and $\bP_V = \V^{k+1}(\V^{k+1})\trans$. Then by some algebraic simplifications, we have
\begin{align*}
    \bGamma_{I^k_{22}\cdot}^{k+1}
    = \X_{\cdot I_{22}^k}\trans (\I - \bP_{A^k})\Y\V^{k+1}
    ~\text{and}~
    \wt{\bGamma}^{k+1}_{I^k_{22}\cdot}
    = \X_{\cdot I_{22}^k}\trans (\I - \bP_{A^k})\Y\bP_V + \X_{\cdot I_{22}^k}\trans\Y(\I - \bP_V).
\end{align*}
By Lemma \ref{lemma:4}, we can obtain
\begin{align}\label{ineq:3:lem:3}
    \bGamma_{I^k_{22}\cdot}^{k+1}
    = \norm{\X_{\cdot I_{22}^k}\trans (\I - \bP_{A^k})\Y\bP_V}_{\op}
    \leq \norm{\wt{\bGamma}^{k+1}_{I_{22}^k\cdot}}_{\op}.
\end{align}
Combining \eqref{ineq:1:lem:3}--\eqref{ineq:3:lem:3} yields
\begin{align*}
    \norm{\B_{A^k_{11}\cdot}^{k+1}}_{\op} + \norm{\bGamma_{A^k_{22}\cdot}^{k+1}}_{\op}
    \leq \sqrt{r} (\norm{\B_{I^k_{11}\cdot}^{k+1}}_{\op} + \norm{\wt{\bGamma}_{I^k_{22}\cdot}^{k+1}}_{\op}).
\end{align*}
Finally by the definition of $\B^{k+1}$ and $\C^{k+1}$, and utilizing Lemma \ref{lemma:4} again, we have $\norm{\B_{I^k_{11}\cdot}^{k+1}}_{\op} = \norm{\C_{I^k_{11}\cdot}^{k+1}}_{\op}$. Thus there is
\begin{align*}
    \norm{\B_{A^k_{11}\cdot}^{k+1}}_{\op} + \norm{\bGamma_{A^k_{22}\cdot}^{k+1}}_{\op}
    \leq \sqrt{r} (\norm{\C_{I^k_{11}\cdot}^{k+1}}_{\op} + \norm{\wt{\bGamma}_{I^k_{22}\cdot}^{k+1}}_{\op}),
\end{align*}
which finishes the proof.

\subsection{Lemma \ref{lemma:5} and its proofs}

\begin{lemma}\label{lemma:5}
    Assume that Conditions \ref{cond:1} and \ref{cond:2} hold. Following the algorithm of MrBeSS, we have
    \begin{align*}
        \norm{\wt{\bGamma}^{k+1}_{A_{22}^k\cdot}}_{\op}
        \leq \norm{\bGamma^{k+1}_{A_{22}^k\cdot}}_{\op} + c h(s),
    \end{align*}
    where $c$ is a positive constant, and $h(s) = \max_{\abs{A}\leq s}~n^{-1}\norm{\X_{\cdot A}\trans\E}_{\op}$.
\end{lemma}

Recall the definition of $\wt{\bGamma}^{k+1}_{A_{22}^k\cdot}$ and $\bGamma^{k+1}_{A_{22}^k\cdot}$, we have
\begin{align*}
    \wt{\bGamma}^{k+1}_{A_{22}^k\cdot}
    = \bGamma^{k+1}_{A_{22}^k\cdot}(\V^{k+1})\trans + \X_{\cdot A_{22}^k}\trans\Y(\I - \bP_V) / n,
\end{align*}
after some simplifications. Here we denote $\bP_V = \V^{k+1}(\V^{k+1})\trans$ for convenience. Utilizing Lemma \ref{lemma:4}, we can get
\begin{align}
    \norm{\wt{\bGamma}^{k+1}_{A_{22}^k\cdot}}_{\op}
     & \leq \norm{\bGamma^{k+1}_{A_{22}^k\cdot}}_{\op}
    + n^{-1}\norm{\X_{\cdot A_{22}^k}\trans\Y(\I - \bP_V)}_{\op}.
    \label{ineq:1:lem:5}
\end{align}

Then we bound the second term in \eqref{ineq:1:lem:5}. Note that there is $\rank{\X_{\cdot A^k}\trans\X\C^*}=r^*$ without loss of the generality. Moreover, we have $\rank{\X\trans\X\C^*}=r^*$. Thus there exists a linear transformer $\T\in\mbR^{\abs{A_{22}^k}\times \abs{A^k}}$ satisfying $\X_{A_{22}^k}\trans\X\C^* = \T\X_{A^k}\trans\X\C^*$, since $\X_{A_{22}^k}\trans\X\C^*$ is a submatrix of the original. It ensures that
\begin{align}
     & n^{-1}\norm{\X_{\cdot A_{22}^k}\trans\X\C^*(\I-\bP_V)}_{\op}
    \leq n^{-1} \norm{\T}_{\op} \norm{\X_{\cdot A^k}\trans\X\C^*(\I-\bP_V)}_{\op}
    \nonumber                                                                                       \\
     & \quad~ \leq C n^{-1} \norm{\X_{\cdot A^k}\trans\X\C^*(\I-\bP_V)}_{\op}, \label{ineq:2:lem:5}
\end{align}
where $C$ is a positive constant, and we use the fact $\norm{\T}_{\op}$ is bounded from above because the columns of $\X$ are normalized.

Next we show the upper bound of $n^{-1} \norm{\X_{\cdot A^k}\trans\X\C^*(\I-\bP_V)}_{\op}$. Utilizing the decomposition in Lemma \ref{lemma:1} again, we have
\begin{align*}
     & \X_{\cdot A^k}\trans\X\C^*(\I-\bP_V)
    = \X_{\cdot A^k}\trans\Y(\I-\bP_V) - \X_{\cdot A^k}\trans\E(\I-\bP_V)
    \nonumber                                                                                                                            \\
     & \quad~ = (\X_{\cdot A^k}\trans\X_{\cdot A^k})^{1/2}\sum_{\ell=1}^q d_{\ell}^{k+1}\u_{\ell}^{k+1}(\v_{\ell}^{k+1})\trans(\I-\bP_V)
    - \X_{\cdot A^k}\trans\E(\I-\bP_V).
\end{align*}
Due to the orthogonality of $\v^{k+1}_{\ell}$ and $\bP_V$, there is
\begin{align*}
    \X_{\cdot A^k}\trans\X\C^*(\I-\bP_V)
    = (\X_{\cdot A^k}\trans\X_{\cdot A^k})^{1/2}\sum_{\ell={r+1}}^q d_{\ell}^{k+1}\u_{\ell}^{k+1}(\v_{\ell}^{k+1})\trans(\I-\bP_V)
    - \X_{\cdot A^k}\trans\E(\I-\bP_V).
\end{align*}
Then following the similar argument in Lemma \ref{lemma:1}, we can obtain
\begin{align*}
     & n^{-1}\norm{\X_{A^k}\trans\X\C^*(\I-\bP)}_{\op}
    \leq n^{-1}d_{r+1}^{k+1}\norm{(\X_{\cdot A^k}\trans\X_{\cdot A^k})^{1/2}}_{\op}
    + n^{-1}\norm{\X_{\cdot A^k}\trans\E}_{\op}
    \nonumber                                                                                                         \\
     & \quad~ \leq n^{-1}\norm{\X_{\cdot A^k}\trans\E}_{\op} \norm{(\X_{\cdot A^k}\trans\X_{\cdot A^k})^{-1/2}}_{\op}
    \norm{(\X_{\cdot A^k}\trans\X_{\cdot A^k})^{1/2}}_{\op}
    + n^{-1}\norm{\X_{\cdot A^k}\trans\E}_{\op},
\end{align*}
where we use Lemma \ref{lemma:4} in the first inequality. Furthermore, under Conditions \ref{cond:1}, \ref{cond:2} and $h(s)=n^{-1}\underset{\abs{A}\leq s}{\max}~\norm{\X_{\cdot A}\trans\E}$, there is
\begin{align}
    n^{-1}\norm{\X_{A^k}\trans\X\C^*(\I-\bP)}_{\op}
    \leq \left[1+c_{-}(s)c_{+}(s)\right] h(s)
    \label{ineq:3:lem:5}
\end{align}
Combining \eqref{ineq:2:lem:5} and \eqref{ineq:3:lem:5}, we have the upper bound as follows
\begin{align}
    n^{-1}\norm{\X_{\cdot A_{22}^k}\trans\X\C^*(\I-\bP_V)}_{\op}
    \leq C^{\prime} h(s), \label{ineq:4:lem:5}
\end{align}
with a positive constant $C^{\prime}$. Moreover, note that there is
\begin{align}
    n^{-1}\norm{\X_{\cdot A_{22}^k}\trans\Y(\I - \bP_V)}_{\op}
    \leq n^{-1}\norm{\X_{\cdot A_{22}^k}\trans\X\C^*(\I-\bP_V)}_{\op}
    + n^{-1}\norm{\X_{\cdot A_{22}^k}\trans\E}_{\op}. \label{ineq:5:lem:5}
\end{align}
Finally, combining \eqref{ineq:1:lem:5}, \eqref{ineq:4:lem:5} and \eqref{ineq:5:lem:5}, we will get the result as follows
\begin{align*}
    \norm{\wt{\bGamma}^{k+1}_{A_{22}^k\cdot}}_{\op}
    \leq \norm{\bGamma^{k+1}_{A_{22}^k\cdot}}_{\op} + c h(s),
\end{align*}
where $c$ is a positive constant. Then it concludes the proof.

\subsection{Lemma \ref{lemma:10} and its proofs}

\begin{lemma}\label{lemma:10}
    Under Conditions \ref{cond:1} and \ref{cond:3}, for any subset $A$ with $\abs{A}\leq s$, the inequality 
    \begin{align*}
        \underset{\abs{A}\leq s}{\max}\,\,n^{-1}\norm{\X_{\cdot A}\trans\E}_{\op} 
    \leq O\left(
        n^{-1/2}\{s + q + s\log(ep/s)\}^{1/2}
    \right),
    \end{align*}
    holds with the probability at least $1 - p^{-c}$, where $c$ is an absolute constant. 
\end{lemma}

Note that the noise matrix $\E$ can be rewritten as $\E = \wt{\E}\bSigma^{1/2}$, where the entries of $\wt{\E}$ are i.i.d from $N(0,1)$. Moreover, there is $\norm{\X_{\cdot A}\trans\E}_{\op} \leq \norm{\X_{\cdot A}\trans\wt{\E}}_{\op}\norm{\bSigma^{1/2}}_{\op}$. Thus together with Condition \ref{cond:3}, the bounded $\norm{\bSigma^{1/2}}_{\op}$, we only need to prove the upper of $n^{-1}\norm{\X_{\cdot A}\trans\wt{\E}}_{\op}$ for an arbitrary $A$. 

Consider the two projection matrices $\bP_A = \X_{\cdot A}(\X_{\cdot A}\trans \X_{\cdot A})^{-1}\X_{\cdot A}\trans$ and $\bP_v = \v\v\trans$ with an arbitrary $\v\in\mbR^q$ and $\norm{\v}_2=1$. Then we have 
\begin{align}\label{ineq:lem:10}
    \max_{\norm{\v}_2=1}\,\, \norm{\bP_A\wt{\E}\bP_v}_{F} 
    \geq \max_{\norm{\v}_2=1}\,\, \norm{\bP_A\wt{\E}\bP_v}_{\op} 
    \geq \frac{1}{c_{+}(s)\sqrt{nc_{-}(s)}}\max_{\norm{\v}_2=1}\,\, \norm{\X_{\cdot A}\trans\wt{\E}\bP_A}_{\op},
\end{align}
where the last inequality is due to Condition \ref{cond:1}. Note that $\underset{\norm{\v}_2=1}{\max}\, \norm{\X_{\cdot A}\trans\wt{\E}\bP_v}_{\op} = \norm{\X_{\cdot A}\trans \wt{\E}}_{\op}$ and thus we mainly prove the upper bound of $R_{A,1}$, where 
\begin{align*}
    R_{A,1} = \underset{\abs{A}\leq s,\norm{\v}_2=1}{\max}\, \norm{\bP_A\wt{\E}\bP_v}_{F}^2.
\end{align*} 
Directly applying Lemma 3 in \citet{she2017selective}, we can get 
\begin{align*}
    P(R_{A,1} > t\sigma^2 + C_1 \sigma^2\{s + q + s\log(ep/s)\}) 
    \leq C_2\exp(-ct),
\end{align*}
where $C_1,C_2$ and $c$ are the absolute constants. Let $t$ be $\log p$ that yields $R_{A,1} \leq O(s + q + s\log(ep/s))$ with the probability at least $1 - p^{-c}$ because of $\log p \leq s\log(ep/s)$ when $s\geq 1$. Together with \eqref{ineq:lem:10} and Condition \ref{cond:3}, we have 
\begin{align*}
    \underset{\abs{A}\leq s}{\max}\,\,n^{-1}\norm{\X_{\cdot A}\trans\E}_{\op} 
    \leq O\left(
        n^{-1/2}\{s + q + s\log(ep/s)\}^{1/2}
    \right)
\end{align*}
holds with the probability at least $1 - p^{-c}$, which finishes the proof. 

\subsection{Lemma \ref{lemma:4} and its proofs}

\begin{lemma}\label{lemma:4}
    Assume that $\V\in\mbR^{q\times r}$ and $\V\trans\V=\I_r$. Then for an arbitrary matrix $\A\in\mbR^{n\times q}$ there is
    \begin{align}
         & \norm{\A\bP}_{\op} \leq \norm{\A}_{\op} \leq \norm{\A\bP}_{\op} + \norm{\A(\I-\bP)}_{\op},
        \label{lemma:4:ineq:1}                                                                        \\
         & \norm{\A\V}_{\op} = \norm{\A\bP}_{\op}, \label{lemma:4:ineq:2}
    \end{align}
    where $\bP = \V\V\trans\in\mbR^{q\times q}$.
\end{lemma}

On the one hand, note that $\A = \A\bP + \A(\I-\bP)$ thus we can get the last inequality in \eqref{lemma:4:ineq:1} immediately by the triangle inequality. Thus we mainly prove the first inequality in \eqref{lemma:4:ineq:1}. For an arbitrary vector $\balpha$, there is
\begin{align*}
    \norm{\A\trans\balpha}_2^2 = \norm{\bP\A\trans\balpha}_2^2 + \norm{(\I-\bP)\A\trans\balpha}_2^2,
\end{align*}
where we use the fact $\bP(\I - \bP) = 0$ and $\bP\trans = \bP$. Denote that $\balpha^* = \underset{\norm{\balpha}_2=1}{\arg\max}~\norm{\bP\A\trans\balpha}_2^2$ then we have
\begin{align*}
    \norm{\A}_{\op}^2
    \geq \norm{\A\trans\balpha^*}_2^2
    = \norm{\bP\A\trans\balpha^*}_2^2 + \norm{(\I-\bP)\A\trans\balpha^*}_2^2
    \geq \norm{\A\bP}_{\op}^2,
\end{align*}
since $\norm{\A}_{\op} = \norm{\A\trans}_{\op}$ and $\norm{\A\bP}_{\op} = \norm{\bP\A\trans}_{\op}$. It concludes the inequality \eqref{lemma:4:ineq:1}.

On the other hand, by the definition of the operator norm, there is
\begin{align*}
    \norm{\A\V}_{\op}
    = \underset{\norm{\balpha}_2=1}{\max}~\norm{\A\V\balpha}_2
    = \underset{\substack{\norm{\balpha}_2=1 \\ \balpha\in\mathcal{V}}}{\max}~ \norm{\A\balpha}_2
    = \underset{\norm{\balpha}_2=1}{\max}~\norm{\A\bP\balpha}_2
    = \norm{\A\bP}_{\op},
\end{align*}
where we use the fact $\V\trans\V=\I_r$ and $\bP$ is the projection matrix onto the space $\mathcal{V}$ spanned by the columns of $\V$. Thus we get the equation \eqref{lemma:4:ineq:2}.

\subsection{Lemma \ref{lemma:7} and its proofs}

\begin{lemma}\label{lemma:7}
    Under Conditions \ref{cond:1} and \ref{cond:2}, there is a positive constant $C$ satisfying
    \begin{align*}
        n^{-1}\abs{\tr{\E\trans(\I - \bP_A)\X\C^*}}
         & \leq C \left(n^{-1}\norm{\E\trans\X_{\cdot J}}_{\op} \vee n^{-1}\norm{\E\trans\X_{\cdot A}}_{\op}\right)
        \cdot r^*\norm{\C^*}_{\op},
        \\
        n^{-1}\tr{\E\trans\bP_A\E}
         & \leq C \abs{A} (n^{-1}\norm{\E\trans\X_{\cdot A}}_{\op})^2,
    \end{align*}
    where $A$ is an arbitrary index set and $J = A^* / A$. Furthermore, denote the $j$th eigenvalue of $n^{-1}\Y\trans\bP_A\Y$ and $n^{-1}\C\strans\X\trans\bP_A\X\C^*$ by $d_j(A)$ and $d_j^*(A)$, respectively. We have
    \begin{align*}
        \abs{d_j(A) - d_j^*(A)}
        \leq C (n^{-1}\norm{\E\trans\X_{\cdot A}}_{\op} \norm{\C^*}_{\op}
        + n^{-2} \norm{\E\trans\X_{\cdot A}}_{\op}^2).
    \end{align*}
    Here $\bP_A = \X_{\cdot A}(\X_{\cdot A}\trans\X_{\cdot A})^{-1} \X_{\cdot A}\trans$ is the projection matrix onto the space $\text{span}(\X_{\cdot A})$.
\end{lemma}

We denote that $J = A^* / A$ for convenience then demonstrate the three inequalities as follows.

\noindent\underline{1. The upper bound of $n^{-1}\abs{\tr{\E\trans(\I-\bP_A)\X\C^*}}$.}

Note that $\X\C^* = \X_{\cdot A}\C^*_{A \cdot} + \X_{\cdot J}\C^*_{J \cdot}$ and $\bP_A$ is the projection matrix onto the space $\text{span}(\X_{\cdot A})$. Thus we have
\begin{align*}
    \E\trans(\I-\bP_A)\X\C^*
    = \E\trans(\I-\bP_A)\X_{\cdot J}\C^*_{J\cdot}
    = \E\trans\X_{\cdot J}\C^*_{J\cdot} - \E\trans\bP_A\X_{\cdot J}\C^*_{J\cdot}.
\end{align*}
Applying Cauchy's inequality to the above inequality, we can obtain
\begin{align*}
     & n^{-1}\abs{\tr{\E\trans(\I-\bP_A)\X\C^*}}
    \leq n^{-1} \abs{\tr{\E\trans\X_{\cdot J}\C^*_{J\cdot}}}
    + n^{-1} \abs{\tr{\E\trans\bP_A\X_{\cdot J}\C^*_{J\cdot}}}
    \\
     & \quad~ \leq n^{-1}\norm{\E\trans\X_{\cdot J}}_{\op} \norm{\C^*}_{*}
    + n^{-1}\norm{\E\trans\X_{\cdot A}}_{\op} \norm{(\X_{\cdot A}\trans\X_{\cdot A})^{-1}\X_{\cdot A}\trans\X_{\cdot J}\C^*_{J\cdot}}_{*},
\end{align*}
where $\norm{\cdot}_{*}$ is the nuclear norm of matrices. By the further simplification, there is
\begin{align*}
     & \quad~ n^{-1}\abs{\tr{\E\trans(\I-\bP_A)\X\C^*}}
    \\
     & \leq n^{-1}r^* \norm{\E\trans\X_{\cdot J}}_{\op} \norm{\C^*}_{\op}
    + n^{-1}r^*\norm{\E\trans\X_{\cdot A}}_{\op} \norm{(\X_{\cdot A}\trans\X_{\cdot A})^{-1}\X_{\cdot A}\trans\X_{\cdot J}\C^*_{J\cdot}}_{\op}
    \\
     & \leq n^{-1}r^* \norm{\E\trans\X_{\cdot J}}_{\op} \norm{\C^*}_{\op}
    + n^{-1}r^*c_{-}^{-1}(s)\theta_{s}\norm{\E\trans\X_{\cdot A}}_{\op}\norm{\C^*_{J\cdot}}_{\op},
\end{align*}
where we use Conditions \ref{cond:1} and \ref{cond:2} in the last inequality, and $s = \max\{\abs{A}, \abs{J}\}$. Then we get that
\begin{align*}
    n^{-1}\abs{\tr{\E\trans(\I - \bP_A)\X\C^*}}
    \leq C \left(n^{-1}\norm{\E\trans\X_{\cdot J}}_{\op} \vee n^{-1}\norm{\E\trans\X_{\cdot A}}_{\op}\right)
    \cdot r^*\norm{\C^*}_{\op},
\end{align*}
with a positive constant $C$.

\noindent\underline{2. The upper bound of $n^{-1}\tr{\E\trans\bP_A\E}$.}

Similar to the above argument, applying Conditions \ref{cond:1} and \ref{cond:2} yields
\begin{align*}
    n^{-1}\tr{\E\trans\bP_A\E}
     & = n^{-1}\norm{\bP_{A}^{1/2}\E}_F^2
    \leq n^{-1}\norm{(\X_{\cdot A}\trans\X_{\cdot A})^{-1}}_{\op}
    \norm{\X_{\cdot A}\trans\E}_F^2
    \\
     & \leq c_{-}^{-1}(s) n^{-2} \norm{\X_{\cdot A}\trans\E}_F^2
    \leq c_{-}^{-1}(s) \abs{A} n^{-2} \norm{\X_{\cdot A}\trans\E}_{\op}^2,
\end{align*}
where the first inequality is due to the definition of the operator norm. And the last inequality is due to the rank of $\X_{\cdot A}\trans\E$ is no more than $\abs{A}$. Then we get
\begin{align*}
    n^{-1}\tr{\E\trans\bP_A\E}
    \leq C \abs{A} (n^{-1}\norm{\E\trans\X_{\cdot A}}_{\op})^2.
\end{align*}

\noindent\underline{3. The upper bound of $\abs{d_j(A) - d_j^*(A)}$.}

Recall that $d_j(A)$ is the $j$th eigenvalue of $n^{-1}\Y\trans\bP_A\Y$, and there is a decomposition as follows
\begin{align*}
    n^{-1}\Y\trans\bP_A\Y
    = n^{-1} \C\strans\X\trans\bP_A\X\C^*
    + n^{-1} \C\strans\X\trans\bP_A\E
    + n^{-1} \E\trans\bP_A\X\C^*
    + n^{-1} \E\trans\bP_A\E.
\end{align*}
Thus by Wely's Theorem, we have
\begin{align*}
    \abs{d_j(A) - d_j^*(A)}
     & \leq 2 n^{-1} \norm{\E\trans\bP_A\X\C^*}_{\op}
    + n^{-1} \norm{\E\trans\bP_A\E}_{\op}
    \\
     & \leq 2n^{-1} \norm{\E\trans\X_{\cdot A}\C^*_{A\cdot}}_{\op}
    + 2n^{-1} \norm{\E\trans\bP_A\X_{\cdot J}\C^*_{J\cdot}}_{\op}
    + n^{-1} \norm{\E\trans\bP_A\E}_{\op},
\end{align*}
where we use the fact $\X\C^* = \X_{\cdot A}\C^*_{A\cdot} + \X_{\cdot J}\C^*_{J\cdot}$ in the last inequality. By a further calculation, we can get
\begin{equation}\label{lem:7:ineq:1}
    \begin{aligned}
         & \quad~ \abs{d_j(A) - d_j^*(A)}
        \\
         & \leq 2n^{-1} \norm{\E\trans\X_{\cdot A}}_{\op} \norm{\C^*_{A\cdot}}_{\op}
        + 2n^{-1} \norm{\E\trans\X_{\cdot A}}_{\op}
        \norm{(\X_{\cdot A}\trans\X_{\cdot A})^{-1}\X_{\cdot A}\trans\X_{\cdot J}\C^*_{J\cdot}}_{\op}
        \\
         & \quad~ + n^{-1} \norm{\E\trans\X_{\cdot A}}_{\op}^2 \norm{(\X_{\cdot A}\trans\X_{\cdot A})^{-1}}_{\op}.
    \end{aligned}
\end{equation}
Hereafter, we can use the similar argument as above. There are
\begin{equation}\label{lem:7:ineq:2}
    \begin{aligned}
         & \norm{(\X_{\cdot A}\trans\X_{\cdot A})^{-1}\X_{\cdot A}\trans\X_{\cdot J}\C^*_{J\cdot}}_{\op}
        \leq c_{-}^{-1}(s) \theta_{s} \norm{\C^*_{J\cdot}}_{\op},
        \\
         & \norm{(\X_{\cdot A}\trans\X_{\cdot A})^{-1}}_{\op} \leq n^{-1} c_{-}^{-1}(s),
    \end{aligned}
\end{equation}
under Conditions \ref{cond:1} and \ref{cond:2}. Combining \eqref{lem:7:ineq:1} and \eqref{lem:7:ineq:2}, we can get
\begin{align*}
     & \abs{d_j(A) - d_j^*(A)}
    \leq 2n^{-1} \norm{\E\trans\X_{\cdot A}}_{\op} \norm{\C^*_{A\cdot}}_{\op}
    \\
     & \quad~ + 2c_{-}^{-1}(s) \theta_{s} n^{-1} \norm{\E\trans\X_{\cdot A}}_{\op}
    \norm{\C^*_{J\cdot}}_{\op}
    + c_{-}^{-1}(s) n^{-2} \norm{\E\trans\X_{\cdot A}}_{\op}^2.
\end{align*}
Thus it ensures that
\begin{align*}
    \abs{d_j(A) - d_j^*(A)}
    \leq C (n^{-1}\norm{\E\trans\X_{\cdot A}}_{\op} \norm{\C^*}_{\op}
    + n^{-2} \norm{\E\trans\X_{\cdot A}}_{\op}^2),
\end{align*}
which concludes the proof.

\subsection{Lemma \ref{lemma:9} and its proofs}

\begin{lemma}\label{lemma:9}
    Assume that Condition \ref{cond:1} holds then we have
    \begin{align*}
        \abs{L(\wt{\C}^{\ora}) - L(\C^*)}
        \leq C \wt{r} \left(n^{-1}\norm{\E\trans\X_{\cdot A^*}}_{\op} +
        n^{-2}\norm{\E\trans\X_{\cdot A^*}}_{\op}^2 \right),
    \end{align*}
    where $\wt{\C}^{\ora}$ is the reduced rank estimator with oracle support set $A^*$. That is
    \begin{align*}
        \wt{\C}^{\ora}_{A^*\cdot} = \underset{\C\in\mbR^{s^*\times q}}{\arg\min}~
        (2n)^{-1}\norm{\Y - \X_{\cdot A^*}\C}_F^2 \quad \text{s.t.}~\rank{\C}=\wt{r} \geq r^*,
    \end{align*}
    where the other rows excluding $A^*$ in $\wt{\C}^{\ora}$ are zero. Specially, the oracle reduced rank estimator $\wh{\C}^{\ora}$ with the support set $A^*$ and $\rank{\wh{\C}^{\ora}}=r^*$ satisfies 
    \begin{align*}
        \abs{L(\wh{\C}^{\ora}) - L(\C^*)}
        \leq C r^* \left(n^{-1}\norm{\E\trans\X_{\cdot A^*}}_{\op} +
        n^{-2}\norm{\E\trans\X_{\cdot A^*}}_{\op}^2 \right). 
    \end{align*}
\end{lemma}

By direct calculations, we have
\begin{align}
    L(\wt{\C}^{\ora}) - L(\C^*)
    = (2n)^{-1}\norm{\X(\wt{\C}^{\ora} - \C^*)}_F^2 + n^{-1}\tr{\E\trans\X(\C^* - \wt{\C}^{\ora})}. \label{lem:9:ineq:5}
\end{align}
On the one hand, utilizing the minimum of $\wt{\C}^{\ora}$, we can obtain
\begin{align}
    (2n)^{-1}\norm{\X(\wt{\C}^{\ora} - \C^*)}_F^2
     & \leq n^{-1}\tr{\E\trans\X(\wt{\C}^{\ora} - \C^*)}
    \nonumber                                                                                                \\
     & \leq n^{-1}\norm{\E\trans\X_{\cdot A^*}}_{\op} \norm{\wt{\C}^{\ora} - \C^*}_{*}, \label{lem:9:ineq:1}
\end{align}
where $\norm{\cdot}_{*}$ is the nuclear norm. Moreover, note that $\rank{\wt{\C}^{\ora} - \C^*} \leq 2\wt{r}$. Thus there is
\begin{align}
    \norm{\wt{\C}^{\ora} - \C^*}_{*}
    \leq \sqrt{2\wt{r}} \norm{\wt{\C}^{\ora} - \C^*}_{F}. \label{lem:9:ineq:2}
\end{align}
In addition, utilizing Condition \ref{cond:1}, we have
\begin{align}
    (2n)^{-1}\norm{\X(\wt{\C}^{\ora} - \C^*)}_F^2
    \geq 2^{-1} c_{-}(2s^*) \norm{\wt{\C}^{\ora} - \C^*}_F^2. \label{lem:9:ineq:3}
\end{align}
Combining these inequalities \eqref{lem:9:ineq:1},\eqref{lem:9:ineq:2} and \eqref{lem:9:ineq:3}, we can get
\begin{align}
    \norm{\wt{\C}^{\ora} - \C^*}_F
    \leq 2^{3/2}c_{-}^{-1}(2s^*) \sqrt{\wt{r}} n^{-1}\norm{\E\trans\X_{\cdot A^*}}_{\op}. \label{lem:9:ineq:4}
\end{align}
Furthermore, combining \eqref{lem:9:ineq:4}, \eqref{lem:9:ineq:1} and \eqref{lem:9:ineq:2}, we have
\begin{align}
    (2n)^{-1}\norm{\X(\wh{\C}^{\ora} - \C^*)}_F^2
    \leq c_1 \wt{r} n^{-1}\norm{\E\trans\X_{\cdot A^*}}_{\op}, \label{lem:9:ineq:7}
\end{align}
with a positive constant $c_1$.

On the other hand, by Cauchy's inequality, there is
\begin{align*}
    n^{-1}\abs{\tr{\E\trans\X(\C^* - \wt{\C}^{\ora})}}
     & \leq n^{-1}\norm{\E\trans\X_{\cdot A^*}}_{\op} \norm{\wt{\C}^{\ora} - \C^*}_{*}
    \nonumber                                                                                     \\
     & \leq n^{-1}\norm{\E\trans\X_{\cdot A^*}}_{\op} \sqrt{2\wt{r}} \norm{\wt{\C}^{\ora} - \C^*}_F.
\end{align*}
Thus together with \eqref{lem:9:ineq:4}, there is
\begin{align}
    n^{-1}\abs{\tr{\E\trans\X(\C^* - \wt{\C}^{\ora})}}
    \leq c_2 \wt{r} (n^{-1}\norm{\E\trans\X_{\cdot A^*}}_{\op})^2. \label{lem:9:ineq:6}
\end{align}
Finally, combining \eqref{lem:9:ineq:5}, \eqref{lem:9:ineq:7} and \eqref{lem:9:ineq:6}, we can obtain
\begin{align*}
    \abs{L(\wt{\C}^{\ora}) - L(\C^*)}
    \leq C \wt{r} \left(n^{-1}\norm{\E\trans\X_{\cdot A^*}}_{\op} +
    n^{-2}\norm{\E\trans\X_{\cdot A^*}}_{\op}^2 \right),
\end{align*}
with a positive constant. It finishes the proof.

\subsection{Lemma \ref{lemma:8} and its proofs}

\begin{lemma}\label{lemma:8}
    Denote by $\wh{\C}$ the reduced-rank estimation restricted in the row support set $A$ and rank $r$, and define the event as follows
    \begin{align*}
        \mathcal{E} = \left\{n^{-1}\left(\norm{\E\trans\X_{\cdot J}}_{\op}
        \vee \norm{\E\trans\X_{\cdot A}}_{\op}\right) \leq C\tau_n(s)\right\},
    \end{align*}
    with $\tau_n(s)=n^{-1/2}\{s+q+s\log(ep/s)\}^{1/2}$, $s=\max\{\abs{A},\abs{J}\}$ and $J = A^* / A$. Assume that Conditions \ref{cond:1}, \ref{cond:2} and the assumption $\tau_n(s)=o(\norm{\C^*}_{\op})$ hold. Then under the event $\mathcal{E}$, there are four cases as follows.

    \noindent Case 1: $A^*\not\subset A$ and $r \geq r^*$.
    \begin{align*}
        \abs{L(\wh{\C}) - L(\C^*) - (2n)^{-1}\norm{(\I - \bP_A)\X\C^*}_F^2}
        \leq
        C s\tau_n(s) \norm{\C^*}_{\op}.
    \end{align*}
    Case 2: $A^*\subset A$ and $r \geq r^*$.
    \begin{align*}
        \abs{L(\wh{\C}) - L(\C^*)}
        \leq
        C s\tau_n(s) \norm{\C^*}_{\op}.
    \end{align*}
    Case 3: $A^*\not\subset A$ and $r < r^*$.
    \begin{align*}
         \abs{L(\wh{\C}) - L(\C^*) - (2n)^{-1}\norm{(\I - \bP_A)\X\C^*}_F^2 - 2^{-1}\sum_{j=r+1}^{r^*} d_j^*(A)}
         \leq
        C s\tau_n(s) \norm{\C^*}_{\op}.
    \end{align*}
    Case 4: $A^*\subset A$ and $r < r^*$.
    \begin{align*}
        \abs{L(\wh{\C}) - L(\C^*) - 2^{-1} \sum_{j=r+1}^{r^*} \sigma_j^*}
        \leq
        C s\tau_n(s) \norm{\C^*}_{\op}.
    \end{align*}
    Here $\bP_A = \X_{\cdot A}(\X_{\cdot A}\trans\X_{\cdot A})^{-1}\X_{\cdot A}\trans$ is the projection matrix onto the space $\text{span}(\X_{\cdot A})$, $r^*$ is the true rank of $\C^*$ and the loss function $L(\C) = (2n)^{-1}\norm{\Y - \X\C}_F^2$. In addition, $d_j^*(A)$ and $\sigma_j^*$ are the $j$th largest eigenvalue of $n^{-1}\C\strans\X\trans\bP_{A}\X\C^*$ and $n^{-1}\C\strans\X\trans\X\C^*$, respectively.
\end{lemma}

Recall the definition of the loss function $L(\C) = (2n)^{-1}\norm{\Y - \X\C}_F^2$. Moreover note that $\wh{\C}$ is the reduced-rank estimation restricted in row support set $A$ and rank $r$ thus we have
\begin{align*}
    \wh{\C}_{A\cdot}
    = \underset{\C\in\mbR^{\abs{A}\times q}}{\arg\min}~(2n)^{-1}\norm{\Y - \X_{\cdot A}\C}_F^2,~\text{s.t.}~\rank{\C}=r,
\end{align*}
where the other rows in $\wh{\C}$ are zeros. Because $\wh{\C}$ is the reduced-rank estimator, directly applying \citet[Theorem 2.2]{Reinsel1998} yields that
\begin{align*}
    L(\wh{\C}) = (2n)^{-1}\tr{\Y\trans\Y} - 2^{-1}\sum_{j=1}^{r}d_j(A),
\end{align*}
where $d_j(A)$ is the $j$th largest eigenvalue of $n^{-1}\Y\trans\X_{\cdot A}(\X\trans_{\cdot A}\X_{\cdot A})^{-1}\X_{\cdot A}\trans\Y$. For convenience, we denote that $\bP_A = \X_{\cdot A}(\X\trans_{\cdot A}\X_{\cdot A})^{-1}\X_{\cdot A}\trans$. It is the projection matrix onto the space $\text{span}(\X_{\cdot A})$. Moreover, note that there is
\begin{align*}
    (2n)^{-1}\tr{\Y\trans\Y}
    = (2n)^{-1}\tr{\C\strans\X\trans\X\C^*} + n^{-1}\tr{\C\strans\X\trans\E} + (2n)^{-1}\tr{\E\trans\E}.
\end{align*}
Therefore, we can obtain the difference between $L(\wh{\C})$ and $L(\C^*)$ as follows
\begin{align}
     & L(\wh{\C}) - L(\C^*)
    = L(\wh{\C}) - (2n)^{-1}\tr{\E\trans\E}
    \nonumber                                          \\
     & \quad~ = (2n)^{-1}\tr{\C\strans\X\trans\X\C^*}
    + n^{-1}\tr{\C\strans\X\trans\E}
    - 2^{-1}\sum_{j=1}^{r}d_j(A). \label{lem:8:eq:1}
\end{align}
In addition, we can rewrite $2^{-1}\sum_{j=1}^{r}d_j(A)$ as
\begin{align}
    2^{-1}\sum_{j=1}^{r}d_j(A)
    = (2n)^{-1}\tr{\Y\trans\bP_A\trans\Y}
    - 2^{-1}\sum_{j=r+1}^{\abs{A}} d_{j}(A), \label{lem:8:eq:2}
\end{align}
because the maximum rank of $\Y\trans\bP_A\trans\Y$ is $\abs{A}$.

Then similar to the above argument and by some simplifications, we have
\begin{align}
    \tr{\Y\trans\bP_{A}\Y}
    = \tr{\C\strans\X\trans\bP_{A}\X\C^*}
    + 2\tr{\E\trans\bP_{A}\X\C^*}
    + \tr{\E\trans\bP_{A}\E}. \label{lem:8:eq:3}
\end{align}
Combining \eqref{lem:8:eq:1}, \eqref{lem:8:eq:2} and \eqref{lem:8:eq:3}, we can obtain
\begin{equation}\label{lem:8:eq:diff}
    \begin{aligned}
        L(\wh{\C}) - L(\C^*)
         & = (2n)^{-1}\norm{(\I-\bP_A)\X\C^*}_F^2
        + n^{-1}\tr{\E\trans(\I - \bP_A)\X\C^*}
        \\
         & \quad~ - (2n)^{-1}\tr{\E\trans\bP_A\E}
        + 2^{-1}\sum_{j=r+1}^{\abs{A}} d_j(A),
    \end{aligned}
\end{equation}
where we use the fact $(\I - \bP_A)^2 = \I - \bP_A$. Then we consider four cases as follows under the event
\begin{align*}
    \mathcal{E} = \left\{n^{-1}\left(\norm{\E\trans\X_{\cdot J}}_{\op}
    \vee \norm{\E\trans\X_{\cdot A}}_{\op}\right) \leq C\tau_n(s)\right\},
\end{align*}
where $\tau_n(s)$ has been defined in the above. 

\bigskip

\noindent\underline{(1) Case 1: $A^*\not\subset A$ and $r \geq r^*$.} Directly applying the equation \eqref{lem:8:eq:diff}, we have
\begin{align*}
     & \quad~ \abs{L(\wh{\C}) - L(\C^*) - (2n)^{-1}\norm{(\I - \bP_A)\X\C^*}_F^2}
    \\
     & \leq n^{-1}\abs{\tr{\E\trans(\I - \bP_A)\X\C^*}}
    + (2n)^{-1}\tr{\E\trans\bP_A\E} + 2^{-1}\sum_{j=r+1}^{\abs{A}} d_j(A).
\end{align*}
Then by Lemma \ref{lemma:7}, the above inequality can be bounded as follows
\begin{align*}
     & \quad~ \abs{L(\wh{\C}) - L(\C^*) - (2n)^{-1}\norm{(\I - \bP_A)\X\C^*}_F^2}
    \\
     & \leq C \Big\{
    \left(n^{-1}\norm{\E\trans\X_{\cdot J}}_{\op} \vee n^{-1}\norm{\E\trans\X_{\cdot A}}_{\op}\right)
    \cdot r^*\norm{\C^*}_{\op}
    \\
     & ~+ \abs{A} (n^{-1}\norm{\E\trans\X_{\cdot A}}_{\op})^2
    + (\abs{A}-r)(n^{-1}\norm{\E\trans\X_{\cdot A}}_{\op} \norm{\C^*}_{\op}
    + n^{-2} \norm{\E\trans\X_{\cdot A}}_{\op}^2)
    \Big\},
\end{align*}
with a positive constant $C$, where we use the fact $d_j^*(A)=0$ with $j>r^*$, and $d_j^*(A)$ is the $j$th eigenvalue of $n^{-1}\C\strans\X\trans\bP_A\X\C^*$. Thus under event $\mathcal{E}$, there is
\begin{align*}
     & \quad~ \abs{L(\wh{\C}) - L(\C^*) - (2n)^{-1}\norm{(\I - \bP_A)\X\C^*}_F^2}
    \\
     & \leq c \left\{
    (s+r^*-r)\norm{\C^*}_{\op} 
    + s\tau_n(s)
    \right\}\tau_n(s),
\end{align*}
with some positive constant $c$. Note that $r^* < r$ thus we can get
\begin{align}
     \abs{L(\wh{\C}) - L(\C^*) - (2n)^{-1}\norm{(\I - \bP_A)\X\C^*}_F^2}
     \leq
    c^{\prime} s\tau_n(s) \norm{\C^*}_{\op} , \label{lem:8:case:1}
\end{align}
under the assumption $\tau_n(s)=o(\norm{\C^*}_{\op})$, where $c^{\prime}$ is a positive constant.

\bigskip

\noindent\underline{(2) Case 2: $A^*\subset A$ and $r \geq r^*$.} Note that $(\I - \bP_A)\X\C^*=\0$ when $A^*\subset A$. Thus directly applying \eqref{lem:8:case:1}, we have
\begin{align}
    \abs{L(\wh{\C}) - L(\C^*)}
    \leq
    C s\tau_n(s) \norm{\C^*}_{\op}, \label{lem:8:case:2}
\end{align}
with a positive constant $C$.

\bigskip

\noindent\underline{(3) Case 3: $A^*\not\subset A$ and $r < r^*$.} Similar to the argument in Case 1, applying the equation \eqref{lem:8:eq:diff} yields
\begin{align*}
     & \abs{L(\wh{\C}) - L(\C^*) - (2n)^{-1}\norm{(\I - \bP_A)\X\C^*}_F^2 - 2^{-1}\sum_{j=r+1}^{r^*} d_j(A)}
    \nonumber                                                                                                       \\
     & \quad~ \leq
    n^{-1}\abs{\tr{\E\trans(\I - \bP_A)\X\C^*}}
    + (2n)^{-1}\tr{\E\trans\bP_A\E} + 2^{-1}\sum_{j=r^*+1}^{\abs{A}} d_j(A),
\end{align*}
Furthermore applying Lemma \ref{lemma:7}, we have
\begin{align*}
     & \abs{L(\wh{\C}) - L(\C^*) - (2n)^{-1}\norm{(\I - \bP_A)\X\C^*}_F^2 - 2^{-1}\sum_{j=r+1}^{r^*} d_j^*(A)}
    \nonumber                                                                                                         \\
     & \quad~ \leq C \Big\{
    \left(n^{-1}\norm{\E\trans\X_{\cdot J}}_{\op} \vee n^{-1}\norm{\E\trans\X_{\cdot A}}_{\op}\right)
    \cdot r^*\norm{\C^*}_{\op}
    \\
     & ~+ \abs{A} (n^{-1}\norm{\E\trans\X_{\cdot A}}_{\op})^2
    + (\abs{A}-r^*)\left(n^{-1}\norm{\E\trans\X_{\cdot A}}_{\op} \norm{\C^*}_{\op}
    + n^{-2} \norm{\E\trans\X_{\cdot A}}_{\op}^2\right)
    \Big\}.
\end{align*}
Then under the event $\mathcal{E}$, there is
\begin{align*}
     & \abs{L(\wh{\C}) - L(\C^*) - (2n)^{-1}\norm{(\I - \bP_A)\X\C^*}_F^2 - 2^{-1}\sum_{j=r+1}^{r^*} d_j^*(A)}
    \\
     & \quad~ \leq
    C \left\{s\norm{\C^*}_{\op} + s\tau_n(s)\right\}\tau_n(s).
\end{align*}
Similar to Case 1, we have
\begin{align}
    \abs{L(\wh{\C}) - L(\C^*) - (2n)^{-1}\norm{(\I - \bP_A)\X\C^*}_F^2 - 2^{-1}\sum_{j=r+1}^{r^*} d_j^*(A)}
    \leq C s \tau_n(s)\norm{\C^*}_{\op}, \label{lem:8:case:3}
\end{align}
under the assumption $\tau_n(s) = o(\norm{\C^*}_{\op})$.

\bigskip

\noindent\underline{(4) Case 4: $A^*\subset A$ and $r < r^*$.} Utilizing $(\I - \bP_A)\X\C^*=\0$ again, there is
\begin{align}
    \abs{L(\wh{\C}) - L(\C^*) - 2^{-1}\sum_{j=r+1}^{r^*} d_j^*(A)}
    \leq C s \tau_n(s)\norm{\C^*}_{\op}, \label{lem:8:case:4}
\end{align}
from the inequality \eqref{lem:8:case:3}. Moreover, note that $\X\C^* = \X_{\cdot A^*}\C^*_{A^*\cdot}$ and $\bP_A\X_{\cdot A^*} = \X_{\cdot A^*}$ when $A^* \subset A$. $d_j^*(A)$ also is the $j$th largest eigenvalue of $n^{-1}\C\strans\X\trans\X\trans\C^*$. We rewrite the eigenvalue as $\sigma^*_j$ for clarity. Therefore, the above four cases conclude the results.